%% file: fermion-to-qubit.tex
\documentclass[a4paper,onecolumn,unpublished]{quantumarticle}
\pdfoutput=1

\input{preamble/packages}

\input{preamble/symbols}

\begin{document}

\title{From Fermions to Qubits: A ZX-Calculus Perspective}
\author{Haytham McDowall-Rose}
\affiliation{Department of Computer Science, University of Oxford, United Kingdom}
\author{Razin A.\@ Shaikh}
\affiliation{Department of Computer Science, University of Oxford, United Kingdom}
\affiliation{Quantinuum, 17 Beaumont Street, Oxford, United Kingdom}
\author{Lia Yeh}
\affiliation{Department of Computer Science, University of Oxford, United Kingdom}
\affiliation{Quantinuum, 17 Beaumont Street, Oxford, United Kingdom}

\maketitle

\begin{abstract}

\subfile{sections/Abstract}

\end{abstract}

\section{Introduction}
\subfile{sections/Introduction}

\section{Preliminaries} \label{sec:preliminaries}

\subfile{sections/Background}

\section{Linear encodings} \label{sec:linear-encodings}
\subfile{sections/encodings}

\section{Ternary tree mappings} \label{sec:ternary-tree}

\subfile{sections/ternary-tree}

\section{Local encodings}
\subfile{sections/local-encodings} \label{sec:local-encodings}

\section{Future work}

\subfile{sections/Conclusion}

\section*{Acknowledgements}
We would like to thank Quanlong Wang, Mitchell Chiew, Aleks Kissinger, and Gabriel Greene-Diniz for helpful discussions and feedback on this work.
RS is funded by the Clarendon Fund Scholarship.
LY is funded by the Google PhD Fellowship.

\bibliographystyle{eptcs}
\bibliography{preamble/references-bibtex}

\appendix

\allowdisplaybreaks
\setlength{\jot}{20pt}

\section{Proofs}

\subfile{sections/Proofs}

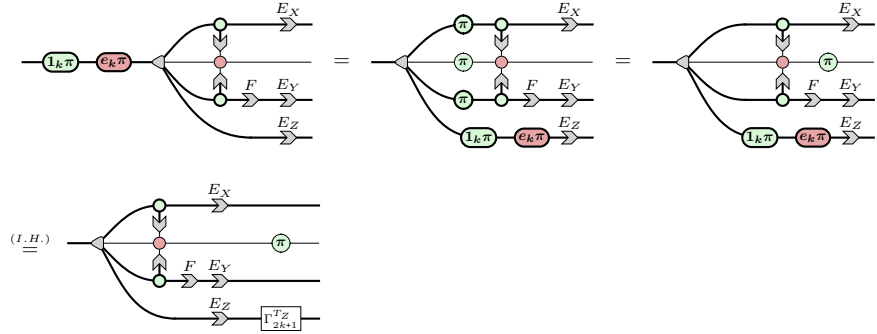
\end{document}

%% file: preamble/packages.tex
\usepackage{bookmark}   
\usepackage[UKenglish]{babel}
\usepackage{colorprofiles}
\usepackage[doipre={doi:}]{uri}   
\hypersetup{
  colorlinks=true,
  citecolor=gray,
  linkcolor=blue,
}

\usepackage{csquotes}

\usepackage{import}
\usepackage{subfiles}

\usepackage{etoolbox}
\apptocmd{\sloppy}{\hbadness 10000\relax}{}{} 

\usepackage{algorithm}
\usepackage{algpseudocode}
\usepackage{mathtools}
\usepackage{physics}
\usepackage{amsfonts}
\usepackage{mathdots}   
\usepackage{mathrsfs}   
\usepackage{bigints}    
\usepackage{xfrac}      
\usepackage{stmaryrd}   
\SetSymbolFont{stmry}{bold}{U}{stmry}{m}{n}   
\usepackage{tabularx}
\usepackage[linewidth=1pt]{mdframed}
\usepackage{microtype}
\usepackage{multicol}
\usepackage{enumitem}
\usepackage{multirow}
\usepackage{amsthm}
\usepackage{thmtools}
\usepackage{thm-restate}

\newtheorem{theorem}{Theorem}
\newtheorem{lemma}{Lemma}
\newtheorem{remark}{Remark}

\newtheorem{proposition}{Proposition}

\usepackage{placeins}

\usepackage{bm}   

\usepackage{blkarray}   

\usepackage{graphicx}
\graphicspath{figures}

\usepackage{tikzfig}
\usepackage{pgfplots} 
\pgfplotsset{compat=newest, compat/show suggested version=false} 
\usetikzlibrary{backgrounds}

\input{preamble/zxw.tikzstyles}
\definecolor{zx_grey}{RGB}{211,211,211}
\definecolor{zx_red}{RGB}{232,165,165}
\definecolor{zx_pink}{RGB}{255, 130, 160}
\definecolor{zx_green}{RGB}{216,248,216}

\usepackage{xtab}
\usepackage{tabu}

\usepackage{float}

\usepackage[capitalise, noabbrev]{cleveref}

%% file: preamble/zxw.tikzstyles
\tikzstyle{boxx}=[draw=black, shape=rectangle, fill=white, minimum size=.95em, inner sep=0.15em, scale=0.85, font={\scriptsize}]
\tikzstyle{gbox}=[boxx, draw=black, shape=rectangle, fill={zx_green}, tikzit fill={rgb,255: red,181; green,215; blue,181}, tikzit shape=rectangle]
\tikzstyle{hbox}=[draw=black, shape=rectangle, fill=yellow, scale=0.6, outer sep=-0.2em]
\tikzstyle{boldhbox}=[hbox, line width=1pt]
\tikzstyle{gn}=[draw=black, shape=circle, fill={zx_green}, inner sep=0.7mm, minimum width=0pt, minimum height=0pt, tikzit fill={rgb,255: red,181; green,215; blue,181}]
\tikzstyle{gn_phase}=[shape=rectangle, fill={zx_green}, draw=black, minimum size=1em, rounded corners=0.44em, inner sep=0.2em, outer sep=-0.2em, scale=0.85, font={\footnotesize\boldmath}, tikzit shape=circle, tikzit fill={rgb,255: red,130; green,188; blue,130}]
\tikzstyle{bold_gn}=[line width=1pt, gn, tikzit fill={rgb,255: red,181; green,215; blue,181}, tikzit draw={rgb,255: red,231; green,165; blue,59}]
\tikzstyle{bold_gn_phase}=[{gn_phase}, line width=1pt, tikzit fill={rgb,255: red,130; green,188; blue,130}, tikzit draw={rgb,255: red,231; green,158; blue,40}]
\tikzstyle{rn}=[gn, fill={zx_red}, draw=black, tikzit fill={rgb,255: red,255; green,165; blue,165}]
\tikzstyle{rn_phase}=[{gn_phase}, fill={zx_red}, draw=black, tikzit fill={rgb,255: red,215; green,96; blue,96}]
\tikzstyle{bold_rn}=[line width=1pt, rn, tikzit fill={rgb,255: red,255; green,165; blue,165}, tikzit draw={rgb,255: red,57; green,190; blue,238}]
\tikzstyle{bold_rn_phase}=[line width=1pt, {rn_phase}, tikzit fill={rgb,255: red,215; green,96; blue,96}, tikzit draw={rgb,255: red,57; green,190; blue,238}]
\tikzstyle{lmat}=[shape=signal, signal to=west, signal from=east, fill={zx_grey}, draw=black, minimum height=6pt, inner sep=.75pt, font={\scriptsize \boldmath}, tikzit fill=gray, tikzit category=GLA, outer sep=-2pt]
\tikzstyle{rmat}=[lmat, shape=signal, signal to=east, signal from=west, tikzit fill=gray, tikzit category=GLA]
\tikzstyle{dmat}=[lmat, shape=signal, signal to=west, signal from=east, tikzit fill=gray, tikzit category=GLA, rotate=90]
\tikzstyle{umat}=[lmat, shape=signal, signal to=east, signal from=west, tikzit fill=gray, tikzit category=GLA, rotate=90]
\tikzstyle{d_split}=[shape=trapezium, fill=white, draw=black, inner sep=0pt, trapezium stretches body, text width=15pt, text height=7pt]
\tikzstyle{d_merge}=[{d_split}, shape=trapezium, draw=black, rotate=180]
\tikzstyle{sd_split}=[shape=trapezium, fill=white, draw=black, inner sep=0pt, trapezium stretches body, text width=10pt, text height=5pt]
\tikzstyle{sd_merge}=[{sd_split}, shape=trapezium, draw=black, rotate=180]
\tikzstyle{ZXdimension}=[font={\tiny}, auto, color=gray]
\tikzstyle{rZWdimension}=[fill={gray!10}, font={\tiny}, draw={gray!30}, rounded rectangle, rounded rectangle east arc=none, text={black!60}, inner sep=1.5pt, anchor=east, tikzit fill=gray]
\tikzstyle{lZWdimension}=[fill={gray!10}, font={\tiny}, draw={gray!30}, rounded rectangle, rounded rectangle west arc=none, text={black!60}, inner sep=1.5pt, anchor=west, tikzit fill=gray]
\tikzstyle{wire label}=[font={\tiny}, auto]
\tikzstyle{control}=[draw=black, shape=circle, fill=black, inner sep=0.5mm]
\tikzstyle{uw}=[shape=isosceles triangle, isosceles triangle stretches=true, fill=black, draw=black, minimum width=0.4cm, minimum height=3mm, inner sep=1pt, shape border rotate=90, outer sep=-0.1em]
\tikzstyle{dw}=[shape=isosceles triangle, isosceles triangle stretches=true, fill=black, draw=black, minimum width=0.4cm, minimum height=3mm, inner sep=1pt, shape border rotate=-90, outer sep=-0.1em]
\tikzstyle{rw}=[shape=isosceles triangle, isosceles triangle stretches=true, fill=black, draw=black, minimum width=0.4cm, minimum height=3mm, inner sep=1pt, shape border rotate=0, outer sep=-0.1em]
\tikzstyle{divide}=[regular polygon, regular polygon sides=3, shape border rotate=90, outer sep=-1mm, draw=black, fill={rgb,255: red,211; green, 211; blue, 211}, inner sep=1.6pt, tikzit category=scal, rounded corners=0.8mm, tikzit fill={rgb,255: red,128; green,0; blue,128}]
\tikzstyle{gather}=[divide, shape border rotate=270]
\tikzstyle{splitup}=[divide, shape border rotate=0]
\tikzstyle{splitdown}=[divide, shape border rotate=180]
\tikzstyle{ut}=[shape=isosceles triangle, isosceles triangle stretches=true, fill=yellow, draw=black, minimum width=0.2cm, minimum height=1.5mm, inner sep=1pt, shape border rotate=90, outer sep=-0.08em]
\tikzstyle{dt}=[shape=isosceles triangle, isosceles triangle stretches=true, fill=yellow, draw=black, minimum width=0.2cm, minimum height=1.5mm, inner sep=1pt, shape border rotate=-90, outer sep=-0.08em]
\tikzstyle{dtbold}=[dt, line width=1pt, fill={rgb,255: red,231; green,231; blue,0}]
\tikzstyle{black dot}=[inner sep=0.5mm, minimum width=0pt, minimum height=0pt, fill=black, draw=black, shape=circle]
\tikzstyle{dot}=[black dot]
\tikzstyle{white dot}=[inner sep=0.5mm, minimum width=0pt, minimum height=0pt, fill=white, draw=black, shape=circle]
\tikzstyle{white phase dot}=[minimum size=3.2mm, font={\footnotesize\boldmath}, shape=rectangle, rounded corners=1.3mm, inner sep=0.8mm, outer sep=-2mm, scale=0.8, draw=black, fill=white]
\tikzstyle{ket}=[minimum size=2mm, font={\tiny\boldmath}, shape=rectangle, rounded corners=0.9mm, inner sep=0.6mm, outer sep=-2mm, scale=0.8, draw=black, fill=black, text=white]
\tikzstyle{box}=[rectangle, fill=white, draw=black, font={\scriptsize}, inner sep=2pt]
\tikzstyle{box-no-outline}=[rectangle, inner sep=2pt]
\tikzstyle{fswap}=[inner sep=0.7mm, minimum width=0pt, minimum height=0pt, draw=purple, shape=circle]
\tikzstyle{compact dash}=[dash pattern=on 2pt off 1pt]
\tikzstyle{W-1-n}=[draw, trapezium, trapezium left angle=70, trapezium right angle=70, fill=black, inner sep=1.8pt]
\tikzstyle{label}=[font={\scriptsize}]
\tikzstyle{ggen}=[fill=white, draw=black, shape=rectangle, rounded corners=1.5mm, line width=0.5pt, tikzit draw=red, tikzit category=scal, inner sep=1.3mm]
\tikzstyle{bgen}=[ggen, fill={rgb,255: red,199; green,199; blue,199}, draw={rgb,255: red,33; green,37; blue,148}, tikzit draw={rgb,255: red,0; green,25; blue,255}, shape=rectangle]
\tikzstyle{scalenone}=[none, font={\footnotesize\boldmath}, scale=0.85]

\tikzstyle{snekedge}=[-, braceedge, decoration={brace, amplitude=0.5mm, raise=-0.3mm}, draw={rgb,255: red,66; green,166; blue,166}]
\tikzstyle{braceedge}=[-, decorate, decoration={brace, amplitude=2mm, raise=-1mm}]
\tikzstyle{dotsedge}=[-, dotted, decoration={brace, amplitude=2mm, raise=-1mm}]
\tikzstyle{boldedge}=[-, line width=1pt, shorten <=-0.17mm, shorten >=-0.17mm, tikzit draw={rgb,255: red,212; green,70; blue,198}]
\tikzstyle{filllayer}=[-, fill={rgb,255: red,250; green,225; blue,200}, opacity=0.5]
\tikzstyle{snekedgeb}=[-, snekedge, line width=1 pt, draw={rgb,255: red,38; green,100; blue,100}]
\tikzstyle{filllayergray}=[-, filllayer, fill={rgb,255: red,150; green,150; blue,150}, draw=white, line width=0 pt]
\tikzstyle{filllayerlight}=[-, filllayergray, fill={rgb,255: red,220; green,220; blue,220}]

%% file: preamble/symbols.tex
\ifdef{\C}
  {\renewcommand{\C}{\mathbb{C}}}
  {\newcommand{\C}{\mathbb{C}}}
\newcommand{\minu}{\texttt{-}}
\newcommand{\plus}{\texttt{+}}

\makeatletter
\newcommand{\oset}[3][1.5ex]{%
  \mathrel{\mathop{#3}\limits^{
    \vbox to#1{\kern-2\ex@
    \hbox{$\scriptstyle#2$}\vss}}}}
\makeatother

\newcommand{\interp}[1]{\left\llbracket#1\right\rrbracket}

\newcommand{\bR}{\begin{color}{red}}
\newcommand{\bB}{\begin{color}{blue}}
\newcommand{\bM}{\begin{color}{magenta}}
\newcommand{\bC}{\begin{color}{cyan}}
\newcommand{\bW}{\begin{color}{white}}
\newcommand{\bBl}{\begin{color}{black}}
\newcommand{\bG}{\begin{color}{green}}
\newcommand{\bY}{\begin{color}{yellow}}
\newcommand{\e}{\end{color}}

\newcommand{\lt}{\ensuremath <}
\newcommand{\gt}{\ensuremath >}

\newcommand{\eqlabel}[1]{\overset{\textsf{\tiny (#1)}}{=}}

%% file: sections/Abstract.tex
Mapping fermionic systems to qubits on a quantum computer is often the first step for algorithms in quantum chemistry and condensed matter physics.
However, it is difficult to reconcile the many different approaches that have been proposed, such as those based on binary matrices, ternary trees,  and stabilizer codes.
This challenge is further exacerbated by the many ways to describe them---transformation of Majorana operators, action on Fock states, encoder circuits, and stabilizers of local encodings---making it challenging to know when the mappings are equivalent.
In this work, we present a graphical framework for fermion-to-qubit mappings that streamlines and unifies various representations through the ZX-calculus.

To start, we present the correspondence between linear encodings of the Fock basis and phase-free ZX-diagrams.
The commutation rules of scalable ZX-calculus allows us to convert the fermionic operators to Pauli operators under any linear encoding.
Next, we give a translation from ternary tree mappings to scalable ZX-diagrams, which not only directly represents the encoder map as a CNOT circuit, but also retains the same structure as the tree.
Consequently, we graphically prove that ternary tree transformations are equivalent to linear encodings, a recent result by Chiew et al.~\cite{chiewTernaryTreeTransformations2024a}.
The scalable ZX representation moreover enables us to construct an algorithm to directly compute the binary matrix for any ternary tree mapping.
Lastly, we present the graphical representation of local fermion-to-qubit encodings.
Its encoder ZX-diagram has the same connectivity as the interaction graph of the fermionic Hamiltonian and also allows us to easily identify stabilizers of the encoding.

%% file: sections/Introduction.tex
Simulating quantum systems, such as molecules in quantum chemistry or materials in condensed matter physics, is one of the most promising applications of quantum computing.
These quantum systems are fermionic in nature and are described in a Fock space.
The first step in any quantum algorithm for studying them is to map the fermionic representation onto qubits.
This mapping is called a \emph{fermion-to-qubit mapping}.
It describes how the fermionic operators, such as creation and annihilation operators, get mapped to qubit operators.
The earliest example of a fermion-to-qubit mapping is the Jordan-Wigner transform, which maps each fermion to a qubit~\cite{jordanUeberPaulischeAequivalenzverbot1928}.
While it is the simplest mapping, the size of operators (i.e.\@ number of qubits the operator act on non-trivially)  scales linearly with the size of the system.
This led to the development of more efficient mappings such as the Bravyi-Kitaev transform~\cite{Bravyi_2002}, which scales logarithmically with the size of the system, and the Sierpinski triangle mapping~\cite{harrison2024sierpinskitrianglefermiontoqubittransform}, which is even more efficient.

Ternary tree mappings~\cite{vlasovCliffordAlgebrasSpin2022, millerBonsaiAlgorithmGrow2023} offer a novel approach to constructing fermion-to-qubit mappings, which can be tailored to specific problems and hardware requirements.
This method shifts the focus from encoded Fock states to encoded Majorana operators, introducing an entirely new class of mappings.
Jiang et al.~\cite{jiangOptimalFermiontoqubitMapping2020} applied this operator-centric approach to obtain a provably optimal fermion-to-qubit mapping with operators of weight $\lceil \log_3 (2n+1)\rceil$.
These techniques have enabled algorithms for finding mappings tailored to hardware constraints such as limited qubit connectivity~\cite{millerBonsaiAlgorithmGrow2023,millerTreespilationArchitectureStateOptimised2024}.

The Jordan-Wigner transform is well suited for one-dimensional Hamiltonians because it preserves the locality of fermionic operators under the mapping to qubits.
However, for Hamiltonians with higher-dimensional interaction graphs, Jordan-Wigner and other unitary mappings produce highly non-local qubit operators.
\emph{Local encodings} address this issue by mapping local fermionic interactions to local qubit operators.
This is typically achieved by introducing ancillary qubits and stabilizers to eliminate non-local terms in the Hamiltonian~\cite{VerstraeteCirac2005local,Setia2019superfast,Whitfield2016local}.
Recent advancements in local encodings have improved locality, circuit depth, and the ratio of qubits to fermionic modes~\cite{Derby2021compact,ChenXu2023fermiontoqubit,OBrien2024ultrafast}.
An experimental demonstration has also shown local encodings outperforming the Jordan-Wigner transform on a trapped-ion quantum computer~\cite{nigmatullin2024experimentalcompact}.

Given the variety of fermion-to-qubit mappings available from different approaches in terms of different representations, it is important to have an intuitive unified framework for understanding and analyzing them.
We propose to build a unified framework for fermion-to-qubit mappings around the ZX-calculus.
The ZX-calculus is a graphical language designed to reason about quantum information represented by diagrams, through an equational rewriting theory~\cite{coeckeInteractingQuantumObservables2008}.
It can be viewed as a tensor network with two types of nodes, called Z and X spiders, and a set of 8 rewrite rules that are sufficient to prove all equalities of qubit linear maps---a property of the calculus called \emph{completeness}.
While the original ZX-calculus was designed for qubits, it has since been generalized to qudits~\cite{poorCompletenessArbitraryFinite2023}, mixed-dimensions~\cite{wangCompletenessQufiniteZXW2024,poorZXcalculusCompleteFiniteDimensional2024}, and infinite-dimensional Hilbert spaces~\cite{shaikhFockedupZX2024}.

The ZX-calculus has found numerous applications in quantum computation including quantum circuit optimization~\cite{debeaudrapFastEffectiveTechniques2020, kissingerReducingTcountZXcalculus2020}, measurement-based quantum computing~\cite{mcelvanneyFlowpreservingZXcalculusRewrite2023, kissingerUniversalMBQCGeneralised2019}, classical simulation~\cite{kissingerClassicalSimulationQuantum2022, Sutcliffe2024ZXCutting}, quantum error correction~\cite{debeaudrapZXCalculusLanguage2020, rodatz2024distancepreserving}, quantum chemistry~\cite{shaikhHowSumExponentiate2022,defeliceLightMatterInteractionZXW2023}, and quantum education~\cite{dundar-coeckeQuantumPicturalismLearning2023}.
Beyond quantum computing, it has also been applied to condensed matter physics~\cite{eastAKLTStatesZXDiagramsDiagrammatic2022, gorantla2024tensornetworksnoninvertiblesymmetries} and quantum field theory~\cite{shaikhCategoricalSemanticsFeynman2022}.
A modification of the ZX-calculus called the ZW-calculus~\cite{coeckeThreeQubitEntanglement2011} has been used to study fermionic quantum computation~\cite{ngDiagrammaticCalculusFermionic2019}.
In this paper, we use the scalable ZX-calculus~\cite{caretteSZXCalculusScalableGraphical2019}, which provides a compact higher-level notation for ZX-diagrams.
This has been used to reason about quantum algorithms~\cite{CaretteOracles2021}, high-level quantum programs~\cite{borgnaEncodingHighlevelQuantum2023}, and transversal gates in quantum error correction~\cite{KissingerScalable2024}.

In this paper, we show that ZX-calculus is a natural framework for studying fermion-to-qubit mappings.
In particular, we identify a one-to-one correspondence between linear encodings of the Fock basis and unitary phase-free ZX-diagrams.
Moreover, product-preserving ternary tree mappings translate elegantly into scalable ZX-diagrams that preserve the original structure while also representing the encoder map.
Since these encoder ZX diagrams are phase-free, we conclude that every ternary tree yields a linear encoding.
Finally, we represent local encodings using Clifford ZX-diagrams.
This form of the encoder isometry mirrors the connectivity of the Hamiltonian interaction graph, and we can directly read off the stabilizers of the encoding from the ZX-diagram.

\subsection{Overview of the paper}
Section~\ref{sec:preliminaries} provides the necessary background on the ZX-calculus and fermion-to-qubit mappings.
In Section~\ref{sec:linear-encodings}, we introduce one of the most well-studied classes of fermion-to-qubit mappings, called linear encodings.
The Jordan-Wigner, Bravyi-Kitaev, and parity encodings are examples of these.
We show that linear encodings correspond to a specific class of ZX-diagrams: unitary phase-free ZX-diagrams.
Representing this in the scalable notation~\cite{caretteSZXCalculusScalableGraphical2019}, we derive the graphical representation of all terms in electronic Hamiltonians for arbitrary linear encodings.

In Section~\ref{sec:ternary-tree}, we show how ternary tree mappings can be represented using the language of the ZX-calculus.
The corresponding scalable ZX-diagrams not only have the same structure as the ternary trees, allowing us to easily read off the Pauli strings, but also provide a graphical representation of the encoder, i.e.\@ the unitary map that the ternary tree represents.
This unifies the operator-centric perspective of ternary tree mappings with that of encoding Fock states.
By pushing the fermionic operators through the encoder, we can obtain the Pauli strings along with their correct sign.
Further, we prove that the encoder for any ternary tree mapping is a phase-free ZX-diagram, and therefore is a linear encoding.
Although this result was recently proven by Chiew et al.~\cite{chiewTernaryTreeTransformations2024a}, we believe our new graphical proof offers a more intuitive understanding of the correspondence.
In addition, the scalable ZX representation yields a novel algorithm to directly compute the binary matrix of the linear encoding from the ternary tree.

In Section~\ref{sec:local-encodings}, we present graphical representations of local encodings.
We begin by demonstrating how the normal forms of scalable ZX-diagrams can compactly describe any local encoding.
Next, we examine two specific examples: E-type and square lattice auxiliary qubit mappings~\cite{{Steudtner2019AQMs}}.
These examples highlight how the ZX-calculus unifies three key perspectives within a single picture: stabilizers, interaction geometry, and the encoder isometry.
These presentations yield insight into the connection between local encodings and quantum error correction; the latter is natural to study through the ZX-calculus, but little work has been done on graphical reasoning for the former.

%% file: sections/Background.tex
\subsection{Fermionic Systems and Fermion-to-Qubit Mappings}
In second quantization, a system of $n$ fermionic modes is characterized in terms of a creation operator $a_i^\dagger$ and an annihilation operator $a_i$ for each mode.
As opposed to bosonic systems, the canonical anti-commutation relations for fermions are $\{a_i,a_j\}=\{a_i^\dagger , a_j^\dagger \} = 0$ and $\{a_i^\dagger , a_j\} = \delta_{ij} \mathbb{I}$.
This defines a basis of all possible occupation numbers $k_i \in \{0,1\}$ called the Fock basis:
\begin{equation}
    |k_0 k_1 ... k_{n-1}) \coloneqq \prod_{i=0}^{n-1} \left(a_i^\dagger \right)^{k_i} |\textsf{vac})
\end{equation}
where $|\textsf{vac})$ is the vacuum state.
In other words, the fermionic Fock space is spanned by the basis $\{\left|\mathbf{f} \right) \, | \, \mathbf{f}\in \mathbb{F}_2^{n}$\}, where $\mathbb{F}_2^{n}$ is the vector space of $n$-dimensional binary vectors.
Here we follow the convention of using curved kets to denote occupation number states and angled kets to denote qubit states.

Instead of creation and annihilation operators, an alternative way to define fermionic systems is in terms of $2n$ unitary, self-adjoint Majorana operators, for $j \in \{0,...,n-1\}$:
\begin{equation}
\gamma_{2j} = a_j^\dagger + a_j \qquad \qquad \gamma_{2j+1} = i(a_j^\dagger - a_j)
\end{equation}
These operators satisfy the canonical anti-commutation relations $\{\gamma_i, \gamma_j\} = 2 \delta_{ij} \mathbb{I}$.

Fermion-to-qubit mappings enable simulating fermionic systems on qubit systems, by mapping the $2^n$-dimensional fermionic Fock space $\mathcal{H}_f$ to the $m$-qubit Hilbert space $(\mathbb{C}^2)^{\otimes m}$.
The mapping is unitary when $m=n$.
The most canonical way to map $n$ fermions to $n$ qubits is to map each fermion occupation number to the same number qubit state, called the Jordan-Wigner transformation: $\left|k_0 k_1 ... k_{n-1}\right) \mapsto \ket{k_0 k_1 ... k_{n-1}}$.
As this transform identifies the Fock basis with the qubit computational basis, we can define other fermion-to-qubit mappings as qubit linear maps with respect to the Jordan-Wigner transform.
Throughout this paper, we use the Jordan-Wigner transform as a reference point.
We describe a mapping $M: \mathcal{H}_{f} \to (\mathbb{C}^2)^{\otimes m}$ by a qubit isometry $U_M$ that maps the Fock basis as follows:
\begin{equation}
    |\mathbf{f}) \ \ \overset{\text{\tiny Jordan-Wigner}}{\longmapsto} \ \ \ket{\mathbf{f}} \ \ \overset{U_M}{\longmapsto} \ \ U_M \ket{\mathbf{f}}
\end{equation}
In this setting, the Jordan-Wigner transform is the identity map, i.e.\@ $U_M = \mathbb{I}$.
The Jordan-Wigner transform maps the Majorana operators to qubit Pauli operators as follows:
\begin{align}
\gamma_{2j} \mapsto X_j \prod_{i=0}^{j-1} Z_i \qquad \qquad \gamma_{2j+1} \mapsto Y_j \prod_{i=0}^{j-1} Z_i
\end{align}
It is easy to check that the anti-commutation relations are preserved under this mapping.

A typical Hamiltonian of a fermionic system is of the following form:
\begin{equation}\label{eq:fermionic-hamiltonian}
    H = \sum_{i,j} h_{ij} a_i^\dagger a_j + \frac{1}{2} \sum_{i,j,k,l} h_{ijkl} a_i^\dagger a_j^\dagger a_k a_l
\end{equation}
where $h_{ij}$ and $h_{ijkl}$ are coefficients of the one-body (or hopping) terms and two-body terms respectively.
In the Jordan-Wigner encoding, the hopping term $a_p^\dagger a_q$ between sites $p < q$ is mapped to
\begin{equation}
A_p^\dagger \otimes \left(\bigotimes_{i=p+1}^{q-1} Z_i\right) \otimes A_q
\end{equation}
where $A_p = (X_p + iY_p)/2$ is the qubit lowering operator.

\subsection{Quantum Graphical Calculi}
\subsubsection{The ZX-calculus}
A ZX-diagram is a labelled undirected graph with two types of nodes, called spiders.
The Z spider is a tensor which is diagonal in the Z basis, and the X spider is a tensor which is diagonal in the X basis.
They are defined respectively as
\begin{align}
    \input{ZXBackground/Z-spider.tikz} & \qquad \overset{\interp{\cdot}}{\longmapsto} \qquad \ket{0}^{\otimes n}\bra{0}^{\otimes m} + e^{i \alpha} \ket{1}^{\otimes n}\bra{1}^{\otimes m}\\
    \input{ZXBackground/X-spider.tikz} & \qquad \overset{\interp{\cdot}}{\longmapsto} \qquad \ket{+}^{\otimes n}\bra{+}^{\otimes m} + e^{i \alpha} \ket{-}^{\otimes n}\bra{-}^{\otimes m}
\end{align}
The label $\alpha \in [0,2\pi)$ is called the phase of the spider.
We omit the label when the phase is $0$.
When all the spiders in a ZX-diagram have phase $0$, we call it a \emph{phase-free ZX-diagram}.
When all the phases are integer multiples of $\pi/2$, we call it a \emph{Clifford ZX-diagram}.
In addition to the spiders, ZX-diagrams may contain identity wires, swaps, cups and caps.
\begin{equation}
    \input{ZXBackground/identity-wire.tikz} \ \overset{\interp{\cdot}}{\longmapsto} \ \sum_{i} \ket{i}\bra{i}
    \quad \qquad
    \input{ZXBackground/swap.tikz} \ \overset{\interp{\cdot}}{\longmapsto} \ \sum_{i,j} \ket{ij}\bra{ji}
    \quad \qquad
    \input{ZXBackground/cup.tikz} \ \overset{\interp{\cdot}}{\longmapsto} \ \sum_{i} \ket{ii}
    \quad \qquad
    \input{ZXBackground/cap.tikz} \ \overset{\interp{\cdot}}{\longmapsto} \ \sum_{i} \bra{ii}
\end{equation}
We can represent the Pauli operators using these spiders:
\begin{equation}
    \input{ZXBackground/paulis.tikz}
\end{equation}
The ZX-calculus consists of ZX-diagrams and a set of rewrite rules that allow us to manipulate them.
The complete set of rewrite rules for the ZX-calculus is shown in Figure~\ref{fig:ZXrules}.
\begin{figure}[h]
    \begin{center}
        \begin{tabular}{|cc|}\hline
            \input{ZXBackground/ZXrule_fuse.tikz} & \input{ZXBackground/ZXrule_picopy.tikz}\\[1em]
            \input{ZXBackground/ZXrule_cc.tikz} & \input{ZXBackground/ZXrule_bialg.tikz}\\[0.7em]
            \input{ZXBackground/ZXrule_id.tikz} & \input{ZXBackground/ZXrule_had.tikz}\\[0.5em]
            \input{ZXBackground/ZXrule_had_decomp.tikz} & \input{ZXBackground/ZXrule_eu.tikz}\\[0.5em]
            \hline
        \end{tabular}
    \end{center}
    \caption{A complete set of rewrite rules~\cite{vilmartNearMinimalAxiomatisationZXCalculus2019}. All the rules hold with their colours swapped.}
    \label{fig:ZXrules}
\end{figure}
These rules hold up to global scalars, which we will often ignore in this paper.
Note that some papers use a different normalization for the X spiders, which makes the rules scalar-exact at the cost of limiting spider phases to $0$ and $\pi$.

\subsubsection{The Scalable ZX-calculus}

In this section, we review the more compact notation of the scalable ZX-calculus~\cite{caretteSZXCalculusScalableGraphical2019}, which will be convenient for reasoning about fermion-to-qubit mappings.
The scalable notation allows us to represent a register of qubits with a single wire.
A bold wire labelled $k$ is a register of $k$ qubits.
\begin{equation}
    \input{ZXBackground/boldwire.tikz}
\end{equation}
To easily regroup the registers of qubits, the divide and gather nodes are defined as follows:
\begin{equation}
    \input{ZXBackground/dividewire.tikz}
    \qquad \qquad
    \input{ZXBackground/gatherwire.tikz}
\end{equation}
Where it is clear from context, we might omit the divide and gather nodes for brevity.
The spiders also have scalable versions, now labelled with a vector of phases $\vec{\alpha} = (\alpha_1, \cdots, \alpha_k) \in \mathbb{R}^k$.
\begin{equation}
    \input{ZXBackground/scalablespider.tikz} \qquad\qquad\qquad \input{ZXBackground/scalableredspider.tikz}
\end{equation}
If the vector $\vec{\alpha}=(\alpha, \cdots, \alpha)$ for some scalar $\alpha$, we may label the bold spider with the scalar $\alpha$.

It is easy to see that all the rewrite rules in Figure~\ref{fig:ZXrules} extend to the scalable ZX-calculus.
In addition, the scalable ZX-calculus introduces a generator called the matrix arrow.
It allows us to compactly represent a bipartite graph of $n$ input $Z$ spiders connected to $m$ output $X$ spiders.
Suppose $A \in \mathbb{F}_2^{m \times n}$ (i.e.\@ an $m \times n$ binary matrix) is the biadjacency matrix of such a graph, where $A_{ij} = 1$ if and only if the $i$th $Z$ spider is connected to the $j$th $X$ spider.
Then the matrix arrow labelled by $A$ is defined as
\begin{equation}\label{eq:matrix-arrow}
    \input{ZXBackground/mat0-def.tikz} \quad \coloneqq \quad \input{ZXBackground/matrixarrowPNF.tikz}
    \qquad \overset{\interp{\cdot}}{\longmapsto} \qquad
    \sum_{\vec{x} \in \mathbb{F}_2^n} \ket{A\vec{x}}\bra{\vec{x}}
\end{equation}
where $\ket{\vec{x}}$, $\vec{x} \in \mathbb{F}_2^n$ is a computational basis state.
We use cups and caps to define the matrix arrow going in the opposite direction:
\begin{equation}
    \input{ZXBackground/mat-transpose-def.tikz}
\end{equation}
As a convention, we omit the matrix arrow label when it is the all-ones matrix.
With the matrix arrows, we have additional rewrite rules.
Composing matrix arrows corresponds to multiplying the matrices:
\begin{equation}
    \input{ZXBackground/mul0.tikz} \quad = \quad \input{ZXBackground/mul1.tikz}
\end{equation}
We can copy matrix arrows through the spiders as shown below:
\begin{equation}
    \input{ZXBackground/copy0.tikz} \quad = \quad \input{ZXBackground/copyAA.tikz}
    \qquad \qquad \qquad
    \input{ZXBackground/copyX0.tikz} \quad = \quad \input{ZXBackground/copyX1.tikz}
\end{equation}
We can commute the scalable Pauli spiders through the matrix arrows.
Let $\vec{u} \in \mathbb{F}_2^n$, then we have
\begin{equation}
    \input{ZXBackground/pivect0.tikz} \quad = \quad \input{ZXBackground/pivect1.tikz}
    \qquad \qquad
    \input{ZXBackground/pivect1-d.tikz} \quad = \quad \input{ZXBackground/pivect0-d.tikz}
\end{equation}
When $A \in \mathbb F_2^{n\times n}$ is invertible,
\begin{equation}
    \input{ZXBackground/matarrowinv1.tikz} \qquad \text{ and } \qquad \input{ZXBackground/matarrowinv2.tikz}
\end{equation}
Finally, we have the rewrites for block matrices:
\begin{equation}
    \input{ZXBackground/raw0.tikz} \ = \ \input{ZXBackground/raw1.tikz}
    \qquad \quad
    \input{ZXBackground/col0.tikz} \ = \ \input{ZXBackground/col1.tikz}
    \qquad \quad
    \input{ZXBackground/block-mat.tikz}
\end{equation}

%% file: figures/ZXBackground/identity-wire.tikz
\begin{tikzpicture}
	\begin{pgfonlayer}{nodelayer}
		\node [style=none] (1) at (0.75, 0) {};
		\node [style=none] (3) at (-0.75, 0) {};
	\end{pgfonlayer}
	\begin{pgfonlayer}{edgelayer}
		\draw (3.center) to (1.center);
	\end{pgfonlayer}
\end{tikzpicture}

%% file: figures/ZXBackground/cup.tikz
\begin{tikzpicture}
	\begin{pgfonlayer}{nodelayer}
		\node [style=none] (0) at (0.25, -0.5) {};
		\node [style=none] (1) at (-0.5, 0) {};
		\node [style=none] (2) at (0.25, 0.5) {};
	\end{pgfonlayer}
	\begin{pgfonlayer}{edgelayer}
		\draw [in=-90, out=180] (0.center) to (1.center);
		\draw [in=90, out=180] (2.center) to (1.center);
	\end{pgfonlayer}
\end{tikzpicture}

%% file: figures/ZXBackground/cap.tikz
\begin{tikzpicture}
	\begin{pgfonlayer}{nodelayer}
		\node [style=none] (0) at (-0.25, -0.5) {};
		\node [style=none] (1) at (0.5, 0) {};
		\node [style=none] (3) at (-0.25, 0.5) {};
	\end{pgfonlayer}
	\begin{pgfonlayer}{edgelayer}
		\draw [in=270, out=0] (0.center) to (1.center);
		\draw [in=-270, out=0] (3.center) to (1.center);
	\end{pgfonlayer}
\end{tikzpicture}

%% file: sections/encodings.tex
We can represent any Fock basis vector $\left|\mathbf{f} \right)$ by its corresponding bitstring $\mathbf{f}$ as a vector in $\mathbb{F}_2^n$.
Given an invertible matrix $A\in \mathbb{F}_2^{n \times n}$, the induced map $\left|\mathbf{f}\right) \mapsto \ket{A \mathbf{f}}$ from the fermionic Fock space to the qubit space is known as a \textit{linear encoding} of the Fock basis.
For example, the Jordan-Wigner encoding corresponds to the identity matrix over $\mathbb{F}_2$.
In this section, we establish the correspondence between linear encodings and phase-free ZX-diagrams, illustrating it with the parity and Bravyi-Kitaev transformations.
We then demonstrate how to compute encoded fermionic operators by pushing them through the encoder, applying this method to all terms in electronic Hamiltonians.

\subsection{Linear Encodings are Phase-Free ZX-Diagrams}
In the ZX-calculus, linear encodings correspond to unitary phase-free ZX-diagrams.
Any such diagram can be simplified to a normal form~\cite{bonchiInteractingBialgebrasAre2014}, which is a matrix arrow shown in Equation~\eqref{eq:matrix-arrow}.
Concretely, the linear encoding given by a matrix $A \in \mathbb{F}_2^{n \times n}$ corresponds to a matrix arrow labelled by $A$.
These unitary phase-free ZX-diagrams can also be rewritten to CNOT circuits~\cite{KissingerWetering2024Book}, i.e.\@ quantum circuits made out of only CNOT gates.

To compute encoded fermionic operators, we employ the technique of \textit{pushing through the encoder}.
This method allows any ZX-diagram to be propagated through any isometry phase-free ZX-diagram~\cite{huangGraphicalCSSCode2023}.
Since we use the Jordan-Wigner encoding as a reference point, we want to apply rewrite rules to can commute a Jordan-Wigner operator past the encoder $E$ for a given encoding.
This gives us the encoded form of the operator because
\begin{equation}
  \input{pushingthroughb.tikz}
\end{equation}
For instance, we can see this in action for creation and annihilation operators in the following proposition.
\begin{restatable}{proposition}{apEgoal}
  \label{prop:apEgoal}
  For a linear encoding $E\in \mathbb{F}_2^{n \times n}$, the encoded fermionic operators $a_p^E$ and $a_p^{E \dagger}$ can be expressed as the following diagrams:
  \begin{equation}
    a_p^E \quad = \quad \scalebox{0.9}{\input{a_pEproof/a_pEgoal.tikz}} \qquad \textnormal{ and } \qquad a_p^{E \dagger} \quad = \quad \scalebox{0.9}{\input{a_pEproof/a_pE-dagger-goal.tikz}}
  \end{equation}
  where $e_p\in \mathbb{F}_2^n$ is the column vector with a $1$ in position $p$ and zeroes elsewhere, and $1_{p-1}=\sum_{q<p}e_p$.
\end{restatable}
\begin{proof}
  We only prove for the annihilation operator $a_p$; the proof for the creation operator is similar.
  The Jordan-Wigner operator for $a_p$ is given by
  \begin{equation}
    a_p \quad  \mapsto \quad \left(\prod_{i=0}^{p-1}Z_i\right) \ket{0_p}\bra{1_p} \quad = \quad \scalebox{0.9}{\input{a_pEproof/a_pJW.tikz}}
  \end{equation}
  We write this Jordan-Wigner operator in the scalable notation and push it through the encoder $E$.
  \[
  \begin{array}{rcl@{\quad}rcl@{\quad}rcl}
  \scalebox{0.9}{\input{a_pEproof/a_pEproof1.tikz}} & = & \scalebox{0.9}{\input{a_pEproof/a_pEproof2.tikz}} & = & \scalebox{0.9}{\input{a_pEproof/a_pEproof3.tikz}} & = \\[3em]
  \scalebox{0.9}{\input{a_pEproof/a_pEproof4.tikz}} & = & \scalebox{0.9}{\input{a_pEproof/a_pEproof5.tikz}} & = & \scalebox{0.9}{\input{a_pEproof/a_pEproof6.tikz}}
  \end{array}
  \]
\end{proof}
We will now look at two classic examples of fermion-to-qubit mappings: the parity and Bravyi-Kitaev encodings.
\subsubsection{The Parity Encoding}
The parity encoding \cite{Seeley_2012} on $N$ qubits is defined by the matrix $E_N$ with $1$s on and below the diagonal, and $0$s elsewhere.
Then, if $N=j+k \text{  for  } 0\leq j,k \leq N$, with $J$ being the appropriately sized matrix of all ones,

$$E_{N}=\begin{bmatrix} 
  E_{j} & \mathbf{0} \\ 
  J  & E_{k} 
\end{bmatrix}$$
We can express this matrix recurrence in the scalable ZX-calculus using matrix arrows. 

\begin{restatable}{proposition}{parityrecurrence}
\label{lem:parityrecurrence}
  In the Scalable ZX-calculus, this recurrence can be expressed as 
  $$\scalebox{0.9}{\input{ParityRecurrenceLHS.tikz}} \quad = \quad \scalebox{0.9}{\input{ParityRecurrenceRHS.tikz}}$$
\end{restatable}
\begin{proof}
  See Appendix~\ref{sec:proofs-linear-encodings}.
\end{proof}
Using this recurrence relation for the encoder circuit, we can derive the form of the annihilation operators under the parity encoding.
\begin{restatable}{proposition}{ParityOperatorShape}
\label{prop:ParityOperatorShape}
  Under the parity encoding , the annihilation operators take the form:
  $$a_p^{Par}=\scalebox{0.9}{\input{ParityOperator.tikz}}$$
\end{restatable}
\begin{proof}
  See Appendix~\ref{sec:proofs-linear-encodings}.
\end{proof}

Note that while the Jordan-Wigner encoding computes the phase factor $(-1)^{\sum_{i=0}^{p-1}n_i}$ by applying Pauli $Z$ gates to the first $p-1$ qubits, the parity transform stores $\sum_{i=0}^{p-1}n_i$ in the $p^{th}$ qubit, and applies a single Pauli $Z$ gate.

\subsubsection{The Bravyi-Kitaev Encoding}
Introduced by Bravyi and Kitaev in \cite{Bravyi_2002}, this encoding allows us to encode fermionic operators on $n$ fermionic sites into qubit operators of weight $O(\log_2(n))$, compared to the weight $O(n)$ required by the Jordan-Wigner and parity encodings.
The encoder is defined recursively by the $\mathbb{F}_2$ matrix $\beta_k$, where 
$$\beta_0= [1], \text{ and for } k\geq 1, \quad  \beta_{k+1}=\begin{bmatrix} 
  \beta_{k} & 0 \\ 
  A & \beta_{k} 
\end{bmatrix}$$
where $A$ is the $(2^k \times 2^k)$ matrix with all ones on its last row, and zeros elsewhere~\cite{Seeley_2012}.
For example,
$$\beta_2= \begin{bsmallmatrix} 
  1 & 0 & 0 & 0 \\ 
  1 & 1 & 0 & 0 \\
  0 & 0 & 1 & 0 \\ 
  1 & 1 & 1 & 1 
\end{bsmallmatrix}, \scalebox{0.8}{\input{beta2example.tikz}} \quad \textnormal{and } \quad
\beta_3= \begin{bsmallmatrix} 
  1 & 0 & 0 & 0 & 0 & 0 & 0 & 0 \\ 
  1 & 1 & 0 & 0 & 0 & 0 & 0 & 0 \\
  0 & 0 & 1 & 0 & 0 & 0 & 0 & 0 \\ 
  1 & 1 & 1 & 1 & 0 & 0 & 0 & 0 \\
  0 & 0 & 0 & 0 & 1 & 0 & 0 & 0 \\ 
  0 & 0 & 0 & 0 & 1 & 1 & 0 & 0 \\
  0 & 0 & 0 & 0 & 0 & 0 & 1 & 0 \\ 
  1 & 1 & 1 & 1 & 1 & 1 & 1 & 1 
\end{bsmallmatrix}, \scalebox{0.8}{\input{beta3example.tikz}} $$

\begin{restatable}{proposition}{bkrecurrence}\label{prop:bkrecurrence}
In the Scalable ZX-calculus, we can express the recurrence relation for the Bravyi-Kitaev encoder (up to dividing and gathering wires) as follows:
\begin{equation}
  \scalebox{0.9}{\input{bkrecurrenceLHS.tikz}} \quad = \quad \scalebox{0.9}{\input{bkrecurrenceRHS.tikz}}
\end{equation}
In other words, the Bravyi-Kitaev encoder for $2^{k+1}$ qubits is obtained by tensoring two $2^k$-qubit encoders and applying a CNOT gate between their respective last qubit.
\end{restatable}
\begin{proof}
  See Appendix~\ref{sec:proofs-linear-encodings}.
\end{proof}
Finally, we can derive the diagrammatic form of the annihilation operators.
\begin{proposition}
Under the Bravyi-Kitaev encoding, the annihilation operators take the form
  $$a_p^{BK} \quad = \quad \scalebox{0.9}{\input{bkoperator.tikz}}$$
\end{proposition}
\begin{proof}
  Follows from Proposition~\ref{prop:apEgoal}.
\end{proof}

\subsection{Electronic Hamiltonians}\label{sec:elec}
To understand the behaviour of individual molecules in chemistry, we should look at Hamiltonians in terms of the fermionic operators.
An electronic system is a system of electrons and nuclei of a molecule. In general, the Hamiltonian of such a system can be expressed in the form
$$H=\sum_{ij} h_{ij} a^\dagger_ia_j+ \frac{1}{2} \sum_{ijkl} h_{ijkl} a^\dagger_ia^\dagger_j a_ka_l$$
where the coefficients $h_{ij}$ and $h_{ijkl}$ are one-electron and two-electron overlap integrals, quantifying the strength of the interactions between the electrons and the nuclei.
In this section, we show how to depict the terms in any electronic Hamiltonian diagrammatically.
For this, we use an extension of the ZX-calculus called the ZXW-calculus~\cite{shaikhHowSumExponentiate2022}.
This extension adds a generator, the W-node:
\begin{equation}
      \scalebox{0.8}{\input{ZXBackground/Wnode-down.tikz}} \quad \overset{\interp{\cdot}}{\longmapsto} \quad \ket{0 \dots 0} \bra{0} + \sum_{i_1 \plus \dots \plus i_n = 1} \ket{i_1 \dots i_n}\bra{1}
\end{equation}
The $W$ node allows us to take linear combinations of (controlled) ZX-diagrams.
A controlled diagram of a linear map $D$ is a diagram $\tilde{D}$ such that
\begin{equation}
  \input{controlledD1.tikz} \qquad \text{ and } \qquad \input{controlledD0.tikz}
\end{equation}
For diagrams $D_1,\cdots D_n$ with controlled diagrams $\tilde{D_1},\cdots \tilde{D_n}$, we can represent (controlled) linear combinations $\sum_{i=1}^{n} \alpha_i D_i$ for scalars $\alpha_1,\cdots, \alpha_n \in \mathbb{C}$ as

\begin{equation}
  \input{controlledlincomb.tikz}
\end{equation}

and we can express the controlled diagram for composition of these diagrams $\prod_{i}D_i$ by 
\begin{equation}
  \input{controlledcomp.tikz}
\end{equation}

Therefore, if we can represent the operators that make up the Hamiltonian as controlled diagrams, we can then we can add and compose them using the ZXW-calculus.

\begin{figure}[ht]
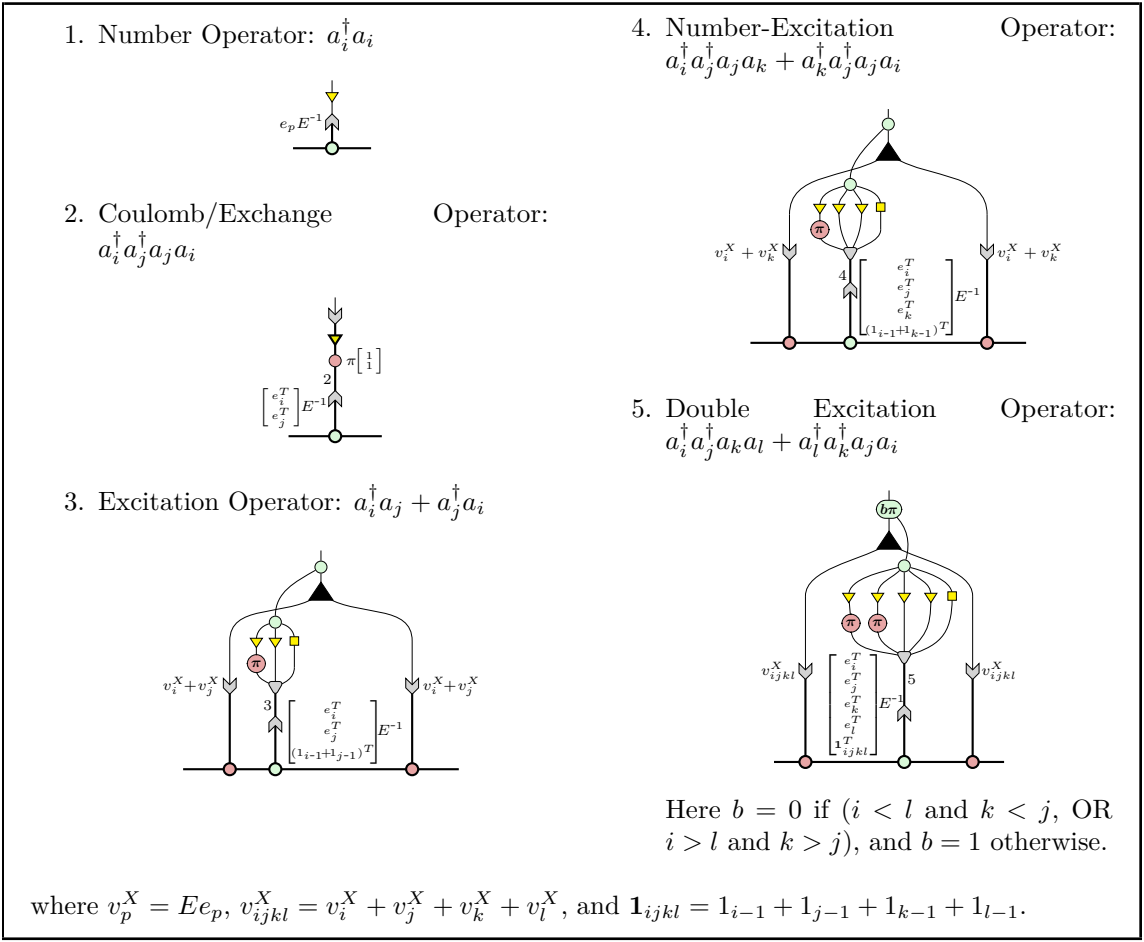

  \begin{mdframed}
    \begin{multicols}{2}
      \begin{enumerate}[topsep=0pt]
        \item Number Operator: $a_i^\dagger a_i$
        \begin{equation*}
          \scalebox{0.8}{\input{numberop.tikz}}
        \end{equation*}
        \item Coulomb/Exchange Operator: $a_i^\dagger a_j^\dagger a_j a_i$
        \begin{equation*}
          \scalebox{0.8}{\input{exchangeop.tikz}}
        \end{equation*}
        \item Excitation Operator: $a_i^\dagger a_j+ a_j^\dagger a_i$
        \begin{equation*}
          \scalebox{0.8}{\input{excitationop.tikz}}
        \end{equation*}\\[0.5em]
        \item Number-Excitation Operator: $a_i^\dagger a_j^\dagger a_j a_k + a_k^\dagger a_j^\dagger a_j a_i$
        \begin{equation*}
          \scalebox{0.8}{\input{numberexop.tikz}}
        \end{equation*}
        \item Double Excitation Operator: $a_i^\dagger a_j^\dagger a_k a_l +a_l^\dagger a_k^\dagger a_j a_i$
        \begin{equation*}
          \scalebox{0.8}{\input{doubleexcop.tikz}}
        \end{equation*}
        Here $b = 0$ if ($i<l$ and $k<j$, OR $i>l$ and $k>j$), and $b = 1$ otherwise.
      \end{enumerate}
    \end{multicols}
    where $v^X_p= Ee_p$, $v^X_{ijkl}=v^X_i+v^X_j+v^X_k+v^X_l$, and $\mathbf{1}_{ijkl}=1_{i-1}+1_{j-1}+1_{k-1}+1_{l-1}$.
  \end{mdframed}
  \caption{Terms in electronic Hamiltonians, and their corresponding controlled diagrams. The derivation of these diagrams is shown in Appendix~\ref{sec:proofs-elec}.}
  \label{fig:elecHamTerms}
\end{figure}

In Figure~\ref{fig:elecHamTerms}, we show controlled diagrams for all types of electronic Hamiltonian terms.
We derive these diagrams for arbitrary linear encodings using the scalable ZX notation.
The derivation of the controlled diagrams is shown in Appendix~\ref{sec:proofs-elec}.
Note that we have used the following shorthand notation in those diagrams:
\begin{equation}
  \scalebox{0.8}{\input{ZXBackground/fliptriangledef.tikz}}
\end{equation}

As an example, we use the representation for electronic Hamiltonian operators to encode the Hamiltonian of the hydrogen molecule on 4 qubits under the Bravyi-Kitaev encoding.
Adding all these terms together and simplifying gives the following.
The coefficients of the Hamiltonian and the simplification steps are given in Appendix~\ref{HydrogenCoeffs}.
\begin{equation}
  \scalebox{0.7}{\input{HydrogenHamiltonian.tikz}}
\end{equation}

%% file: sections/ternary-tree.tex
In this section, we first review ternary tree mappings and then present a translation between ternary tree mappings and phase-free ZX-diagrams.
We verify correctness of this translation by proving that pushing Majorana operators through the encoder results in the same Pauli strings as in the ternary tree formalism.
As a consequence, we show that every ternary tree gives rise to a mapping that is a linear encoding of Fock states.

\subsection{Ternary trees based mappings}
A ternary tree is a labelled ordered tree graph where each vertex has at most three children.
An $n$-vertex ternary tree has nodes labelled by $0, \ldots, n-1$.
We orient the tree left-to-right, with the root at the leftmost node, to match the orientation of ZX-diagrams in this paper.
To obtain a set of anti-commuting Pauli operators, we extend the ternary tree by adding unlabelled leaf nodes such that each node has exactly three children.
We call the three branches---top, middle, and bottom---of each node the $X$, $Y$, and $Z$ branches respectively.

The vertex labels of a ternary tree represent qubit indices.
For any ternary tree, we can associate a set of anti-commuting Pauli strings by listing each path from root to leaves where the $X$, $Y$, and $Z$ branches at node $i$ correspond to the Pauli operators $X_i$, $Y_i$, and $Z_i$ respectively.
(See Figure~\ref{fig:ternary-tree-example} for an example.)
For an $n$-vertex ternary tree, we obtain $2n+1$ Pauli strings.
To obtain a ternary tree mapping, we ignore one of the Pauli strings (usually the all $Z$s Pauli string) and assign the remaining $2n$ Pauli strings to the $2n$ Majorana operators.
In general, the mappings obtained this way may not be product-preserving, i.e.\@ the Fock state $|\mathbf{n})$ may not be mapped to a product of computational basis states $\ket{\mathbf{n'}}$.
However, Miller et al.~\cite{millerBonsaiAlgorithmGrow2023} gave a scheme for pairing the Majorana operators that always results in a product-preserving mapping; in fact, it also preserves the vacuum state (i.e.\@ $|0) \mapsto \ket{0})$.
Chiew et al.~\cite{chiewTernaryTreeTransformations2024a} showed that this product-preserving mapping is unique for any given ternary tree, up to symmetries such as fermionic braids and Pauli relabelling.

\subsection{Translating ternary tree mappings to ZX-diagrams}

We now show how to translate ternary trees to ZX-diagrams, and prove its correctness in Theorem~\ref{thm:ternary-tree-correctness}.

The node labels of ternary trees represent qubit indices.
To obtain a clean translation to ZX-diagrams, we fix the order of qubit labels in ternary trees.
For an $n$-vertex ternary tree, we first label the $X$ subtree of the root node with qubits $0, \ldots, a-1$; the root node with qubit $a$; the $Y$ subtree of the root node with qubits $a+1, \ldots, a+b$; and the $Z$ subtree of the root node with qubits $a+b+1, \ldots, n-1$.
We then recursively label the subtrees of each node in the same manner.
This ordering prevents swaps at the end of the ZX-diagram.
Alternatively, to choose an arbitrary labelling of ternary tree nodes, we can start with the above ordering and then apply a permutation to obtain the desired labelling.

\begin{figure}[t]
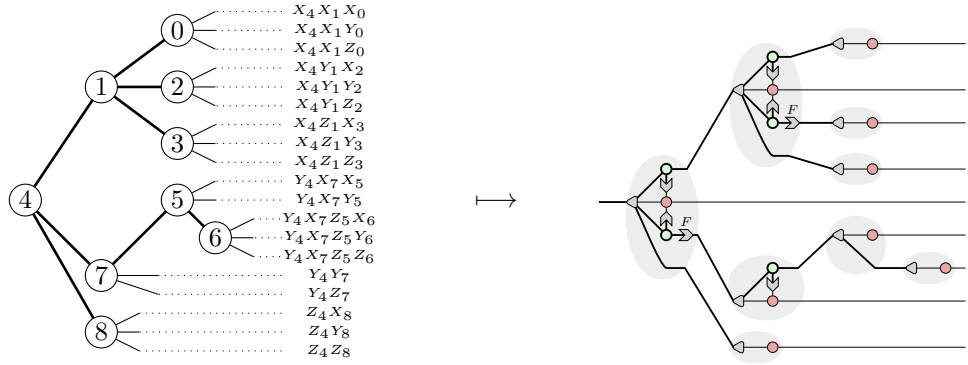

    \centering
    \input{TernaryTree/TTexample2.tikz} \qquad $\longmapsto$ \qquad  \scalebox{0.7}{\input{TernaryTree/TTexamplezx2.tikz}}
    \caption{Left: An example ternary tree with its Pauli strings. Right: Its translated encoder ZX-diagram.}
    \label{fig:ternary-tree-example}
\end{figure}

We can translate any ternary tree mapping to a ZX-diagram by replacing each node in the tree with the following diagram, while keeping the same connectivity as the ternary tree.
\[
    \scalebox{0.7}{\input{TernaryTree/ternary-tree-to-z-x-node.tikz}}
\]
Here, $F$ is a matrix with ones on the anti-diagonal and zeros elsewhere.
Moreover, $a$, $b$, and $c$ are the numbers of descendants of node $i$ down its $X$, $Y$ and $Z$ branches respectively, as well as the numbers of qubits in their scalable ZX wires.
An example of this translation is shown in Figure~\ref{fig:ternary-tree-example}.
For this example ternary tree, we simplify its ZX-diagram to get
\begin{align*}
 \scalebox{0.6}{\input{TernaryTree/TTexamplezx2.tikz}}=\scalebox{0.7}{\input{TernaryTree/TTexample2simp1.tikz}}
= \scalebox{0.7}{\input{TernaryTree/TTexample2simp2.tikz}}=\scalebox{0.7}{\input{TernaryTree/TTexample2simp3.tikz}}
\end{align*}

Miller et al.~\cite{millerBonsaiAlgorithmGrow2023} showed that the classic fermion-to-qubit mappings of Jordan-Wigner, parity and Bravyi-Kitaev can be expressed as ternary trees.
We translate the ternary trees to ZX encoder diagrams and verify that they match the phase-free ZX-diagrams given in Section~\ref{sec:linear-encodings}.
\begin{enumerate}
    \item Jordan-Wigner transform.
    $$\scalebox{0.7}{\input{TernaryTree/JWTree.tikz}} \qquad \longmapsto \qquad \scalebox{0.7}{\input{TernaryTree/JWTernaryTree.tikz}} \qquad  = \qquad \scalebox{0.7}{\input{TernaryTree/JWTernaryTree2.tikz}}$$
    \item Parity transform.
    $$\scalebox{0.7}{\input{TernaryTree/ParityTree.tikz}} \qquad \longmapsto \qquad \scalebox{0.7}{\input{TernaryTree/ParityTernaryTree.tikz}} \qquad  = \qquad  \scalebox{0.7}{\input{TernaryTree/ParityTernaryTree2.tikz}}$$
    \item Bravyi-Kitaev transform.
    \begin{align*}
    \scalebox{0.7}{\input{TernaryTree/BKTree.tikz}} \quad \longmapsto &\quad \scalebox{0.7}{\input{TernaryTree/BKTernaryTree.tikz}} \quad  =  \quad \scalebox{0.7}{\input{TernaryTree/BKTernaryTree2.tikz}}\\
    &= \quad \scalebox{0.7}{\input{TernaryTree/BKTernaryTree3.tikz}} \quad = \quad \scalebox{0.7}{\input{TernaryTree/BKTernaryTree4.tikz}} \quad = \quad \scalebox{0.7}{\input{TernaryTree/BKTernaryTree5.tikz}}
    \end{align*}
    The final diagram is equal to the matrix arrow of $\beta_3$ shown in Section~\ref{sec:linear-encodings}.
\end{enumerate}

\subsection{Ternary trees yield linear encodings}

In this section, we prove that every ternary tree gives rise to a linear encoding.
We gave a translation from ternary tree mappings to unitary phase-free ZX-diagrams in the previous subsection.
Since unitary phase-free ZX-diagrams correspond to linear encodings, all that is left to do is to show that the translation of ternary tree mappings to ZX-diagrams is indeed correct.
To do this, we will prove that pushing the Jordan-Wigner Majorana strings through the encoder gives us all the Pauli strings generated by the ternary tree.
\begin{restatable}{theorem}{ternaryTreeCorrectness}\label{thm:ternary-tree-correctness}
    Translating the ternary tree $T$ to a ZX-diagram by replacing each node in the tree as follows, gives the unitary map corresponding to the ternary tree mapping.
    \[
        \scalebox{0.7}{\input{TernaryTree/ternary-tree-to-z-x-node.tikz}}
    \]
    where $a$, $b$ and $c$ are the numbers of descendants of $i$ down its $X$, $Y$ and $Z$ branches respectively.
\end{restatable}
\begin{proof}
    To prove the correctness, we push the Jordan-Wigner majorana strings through the encoder diagram and show that we obtain all the Pauli strings correctly.
    We only need to verify that we get the correct set of Pauli strings because there is a unique product-preserving mapping for the set of ternary tree Pauli strings, as proved by Chiew et al.~\cite{chiewTernaryTreeTransformations2024a}.

    We prove this by induction on the height of the ternary tree $T$.
    The base case is when $T$ is a leaf node.
    The encoder diagram is an identity wire on a single qubit.
    The two Majorana operators are Pauli $X$ and $Y$ on that qubit.
    It trivially follows that the Jordan-Wigner Majorana strings are mapped to the correct Pauli strings.

    Next, suppose that the proposition holds for ternary trees of height less than $h$.
    Now consider a ternary tree $T$ of height $h$, which has subtrees $T_X$, $T_Y$, and $T_Z$ as shown below.
    \[
        \scalebox{0.9}{\input{TernaryTree/TernaryTreeSubtrees.tikz}}
    \]
    We assume that the qubits are ordered according to the convention described in the main text: first the qubits corresponding to the $X$ subtree, then the current node $i$, and finally the qubits corresponding to the $Y$ and $Z$ subtrees respectively.
    We can write down the set of Pauli strings based on the Pauli strings of the subtrees.
    For all Pauli strings $P_X$, $P_Y$, and $P_Z$ in the subtrees $T_X$, $T_Y$, and $T_Z$, we have the Pauli strings $X_i P_X$, $Y_i P_Y$, and $Z_i P_Z$ respectively.
    In addition, the previously ignored all $Z$ strings of the $X$ and $Y$ subtrees will be composed with $X_i$ and $Y_i$ respectively.

    Now we push all the Jordan-Wigner Majorana strings through the encoder diagram.
    We do this case by case; let the number of nodes in the $X$, $Y$, and $Z$ subtrees be $a$, $b$, and $c$ respectively.
    \begin{enumerate}
        \item Case: $k \lt a$\\
        First we push the majorana operator $\Gamma_{2k+1}$. The proof for $\Gamma_{2k}$ is analogous.
        \begin{equation*}
            \scalebox{0.9}{\input{TernaryTree/proof-X.tikz}}
        \end{equation*}
    \end{enumerate}
    The rest of the cases are similar and proved in Appendix~\ref{sec:ternary-proofs}.
\end{proof}

We can now derive the encoding matrix $E_T$ for the ternary tree mapping.
Chiew et al.~\cite{chiewTernaryTreeTransformations2024a} also described a method of constructing the encoding matrix $E_T$ for a ternary tree $T$.
However, it requires first applying the pairing algorithm to obtain the Pauli strings for the Majorana operators.
In contrast, we now describe a simple recursive algorithm to construct $E_T$ directly from the ternary tree $T$.
It is described in Algorithm~\ref{alg:ternary-tree-encoder}.
\begin{algorithm}
    \caption{Constructing the encoding matrix $E_T$ for a ternary tree $T$}
    \label{alg:ternary-tree-encoder}
    \begin{algorithmic}
        \Function{encoderMatrix}{Tree $T$}
        \If{$T$ is a leaf node}
        \State \Return $\begin{bmatrix} 1 \end{bmatrix}$
        \EndIf
        \State $E_X \ \ \, \gets \text{encoderMatrix}(\text{X subtree of T})$
        \State $E_Y \ \ \, \gets \text{encoderMatrix}(\text{Y subtree of T})$
        \State $E_Y^{\text{flip}} \gets \text{reverse the columns of } E_Y$
        \State $E_Z \ \ \, \gets \text{encoderMatrix}(\text{Z subtree of T})$
        \State \Return $ \begin{bmatrix}
                        E_X            & 0 & 0              & 0 \\
                        \vec{1}^{\top} & 1 & \vec{1}^{\top} & 0 \\
                        0              & 0 & E_Y^{\text{flip}}            & 0 \\
                        0              & 0 & 0              & E_Z \end{bmatrix}$
        \EndFunction
    \end{algorithmic}
\end{algorithm}
This algorithm is derived using the scalable ZX-calculus rules in Theorem~\ref{thm:ternary-tree-encoder}.
\begin{theorem}\label{thm:ternary-tree-encoder}
    Ternary trees yield linear encodings.
    The encoder diagram for the ternary tree mapping can be reduced to phase-free ZX normal form.
    Algorithm~\ref{alg:ternary-tree-encoder} constructs the encoding matrix $E_T$ for the ternary tree $T$.
\end{theorem}
\begin{proof}
    We prove this by induction on the height of the ternary tree $T$.
    The base case is when $T$ is a leaf node.
    In this case, the encoder diagram is a single identity wire corresponding to the encoder matrix $\begin{bmatrix} 1 \end{bmatrix}$.
    Next, we suppose that the theorem holds for ternary trees of height less than $h$.
    We will show that it holds for a ternary tree of height $h$.
    Suppose the encoder matrix for subtrees along the $X$, $Y$, and $Z$ branches are $E_X$, $E_Y$, and $E_Z$ respectively.
    Then the encoder diagram for the ternary tree $T$ is given by the following diagram, which we can reduce to phase-free ZX normal form, i.e.\@ a matrix arrow.
    \begin{equation}
        \input{TernaryTree/encoder-matrix.tikz}
    \end{equation}
    Note that $F$ is the matrix of all ones on the anti-diagonal, and multiplying $E_Y$ by $F$ reverses the columns of $E_Y$.
\end{proof}

%% file: sections/local-encodings.tex
Due to the fermionic anti-commutation relations, even local interactions---such as nearest-neighbor hopping terms---can become non-local when mapped to local systems like qubits.
The geometric arrangement of fermions in a Hamiltonian can be represented by an interaction graph, where vertices correspond to fermionic sites and edges denote interactions between them; common examples include 1D chains and 2D or 3D lattices.
The purpose of local encodings is to preserve locality by mapping local fermionic interactions to local qubit operators.

Local encodings achieve this by introducing ancillary qubits, now making the mapping an isometry.
A set of commuting stabilizers is then added to eliminate the non-local components of the qubit operators in the encoded fermionic Hamiltonian.
While local encodings are often described in terms of stabilizers or auxiliary Hamiltonians, we demonstrate that expressing these stabilizers using ZX-diagrams naturally links them to the encoder picture developed in previous sections.
Moreover, ZX encoder diagrams retain the geometric structure of the problem Hamiltonian, providing clear visual insight into its interaction graph.

Instead of unitaries in the phase-free fragment of the ZX-calculus, we need to consider isometries in the stabilizer fragment of the ZX-calculus.
This approach, while studied extensively in the context of quantum error correction~\cite{kissingerPhasefreeZXDiagrams2022,huangGraphicalCSSCode2023}, has not been explored for fermion-to-qubit mappings.
Unlike error correcting codes, where the goal is to maximize the code distance (i.e. weight of the shortest logical operator), local encodings prioritize preserving the locality of encoded operators.

Linear encodings are unitary phase-free ZX-diagrams, represented by matrix arrows in the scalable ZX-calculus.
To reason about local encodings, we can generalize to any isometry phase-free ZX-diagram, which corresponds to encoders of CSS codes~\cite{KissingerScalable2024} and takes the form:
\begin{equation}
    \scalebox{1}{\input{LocalEncodings/CSS-scalable-encoder.tikz}}
\end{equation}
where $A,B$ are binary matrices.
More broadly, any isometry in the stabilizer ZX-calculus (which corresponds to encoders of stabilizer codes) can be represented in Graph-States-with-Local-Cliffords form~\cite{backensZXcalculusCompleteStabilizer2014}. In the scalable ZX-calculus, we can represent this as:
\begin{equation}
    \scalebox{1}{\input{LocalEncodings/Stabilizer-scalable-encoder.tikz}}
\end{equation}
where $M_1^{\text{inc}}$ and $M_2^{\text{inc}}$ are incidence matrices of the graphs given by the biadjacency matrices $M_1$ and $M_2$ respectively, and LC denotes a local Clifford operation:
\begin{equation}
    \scalebox{1}{\input{LocalEncodings/local-cliffords-def.tikz}} \quad \text{where } \vec{a},\vec{b},\vec{c} \text{ are vectors of integers.} 
\end{equation}
These generalized representations for local encodings allow us to rederive the graphical form of the common Hamiltonian terms, analogous to Section~\ref{sec:elec}.

In the rest of this section, we build up to a ZX-calculus perspective on local encodings.
To build intuition, we first examine the not-fully-local E-type auxiliary qubit mappings~\cite{Steudtner2019AQMs} to demonstrate the essential concepts in a simpler setting.
We then present a fully local encoding, the square lattice auxiliary qubit mapping~\cite{Steudtner2019AQMs}.
This example highlights how ZX-diagrams for local encodings unifies three key perspective within a single picture: stabilizers, interaction geometry, and the encoder isometry.
Other, more efficient mappings can be analyzed in a similar manner, as well as their constructions and classifications, but we leave this exploration for future work.

\subsection{E-type Auxiliary Qubit Mapping}
Consider a grid of fermionic sites, where the interaction graph is a ($L \times L$) square lattice.
The Jordan-Wigner transformation is the most commonly used encoding for fermionic systems.
The Jordan-Wigner transform is a 1D chain of $L^2$ qubits which can be assigned to the sites of the square lattice in any order; the choice of order affects the optimality of the encoding on 2D nearest neighbor qubit architectures, with the optimal solution (without ancillae) presented in Ref.~\cite{chiew2023JWoptimal}.
It has a \emph{connectivity graph} that is a 1D chain of qubits: Hopping terms between qubits that are not adjacent in this 1D connectivity graph necessarily act on all qubits in between the interacting qubits.
However, this is inefficient for fermionic Hamiltonians where the interaction graph is natively 2D or higher.
This brings us to local encodings, which are fermion-to-qubit encodings that can preserve locality of the fermionic Hamiltonian interaction graph.

\begin{figure}[ht]
    \centering
    \scalebox{1.1}{\input{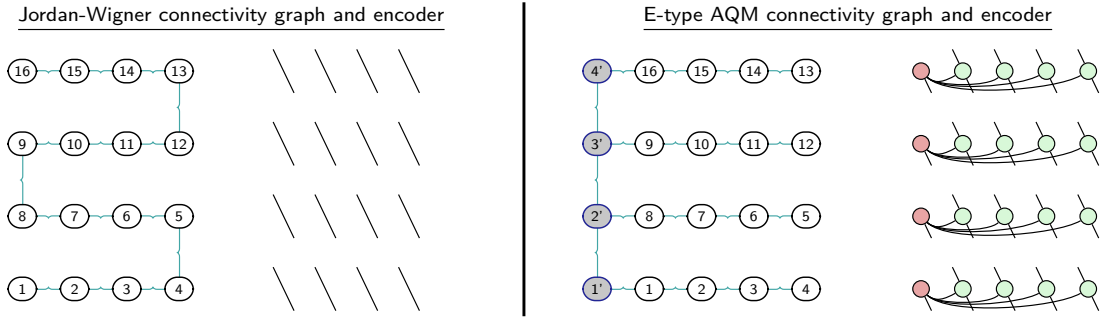}}
    \caption{Left: We see from a possible connectivity graph for the Jordan-Wigner transform, that fermions interacting in a 2D lattice are mapped non-locally to qubits in a 1D chain, even if the qubits themselves are in an architecture that supports 2D nearest-neighbor interactions. Right: Relative to the Jordan-Wigner transform whose encoder is the identity on every qubit, the E-type AQM has an encoder diagram that is a CSS code consisting of adding an ancilla qubit per row computes the parity (i.e. Z...Z stabilizer) of all qubits in that row.}\label{fig:JWvsEencoders}
\end{figure}

While not fully local, the E-type auxiliary qubit mapping (AQM) of Ref.~\cite{Steudtner2019AQMs} improves upon the 2D non-locality of the Jordan-Wigner transform.
This is due to the addition of an extra column of $L$ ancilla qubits, through which the Pauli strings can be routed as a shortcut.
This encoder is presented in the phase-free ZX-calculus in Figure~\ref{fig:JWvsEencoders}.
Throughout this section, we denote inputs to encoder diagrams as wires entering nodes of the ZX-diagram from the top-left, and outputs as leaving from the bottom-right.

\begin{figure}[ht]
    \centering
    \scalebox{1}{\input{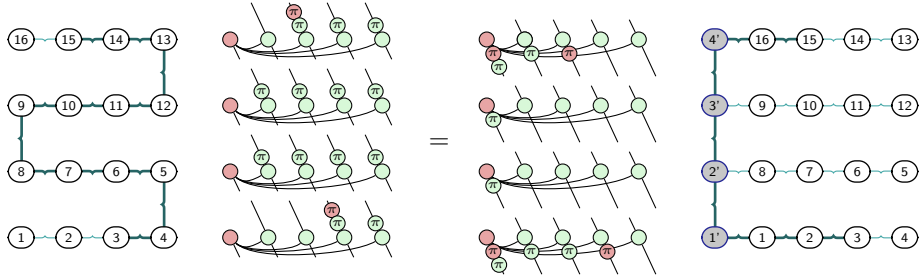}}
    \caption{Consider a hopping term between fermionic sites 3 and 15, which after Jordan-Wigner transform is a weight 13 operator $YZ^{\otimes 8}Y$ on qubits 3 to 15. Pushing this through the E-type AQM encoder, its weight is reduced to 9 by `shortcutting' through the ancillae qubits via the bolded edges of the E-shaped connectivity graph.}\label{fig:JWvsEhopex}
\end{figure}

To derive hopping term operators under the E-type AQM encoding, which are inputs to the encoder of the form $XZ...ZX$ or $YZ...ZY$, we can push them through our diagram for the encoder as shown in Figure~\ref{fig:JWvsEhopex}.

\subsection{Square Lattice Auxiliary Qubit Mapping}
\begin{figure}[ht]
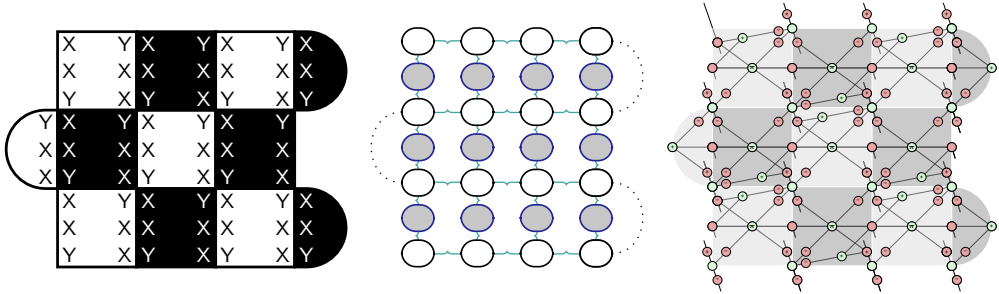

    \centering
    \hspace{-3em}
    \scalebox{2.6}{\input{LocalEncodings/sqAQMtiling.tikz}}
    \hspace{-5.1em}
    \scalebox{1.25}{\input{LocalEncodings/sqAQMgraph4x4.tikz}}
    \hspace{-2em}
    \scalebox{0.6}{\input{LocalEncodings/sqAQMplaquette4x4.tikz}}
    \caption{Left: Tiling of each plaquette stabilizer for the 4 by 4 square lattice AQM from Ref.~\cite{Steudtner2019AQMs}. Middle: Its connectivity graph. Right: Its encoder ZX-diagram. The $\pm$ labels are shorthand for $\pm \frac{\pi}{2}$.}\label{fig:sqAQM}
\end{figure}

In the following, we present an encoder diagram for the square lattice AQM also from Ref.~\cite{Steudtner2019AQMs}.
This mapping interleaves each row with a row of ancilla qubits; according to its connectivity graph in Figure~\ref{fig:sqAQM}, the weight of encoded hopping terms is a function of the Manhattan distance between sites. This is achieved through stabilizers acting on six-qubit plaquettes: four data (non-ancillary) qubits and a pair of same-row adjacent ancilla qubits. Stabilizers are also measured for the boundaries where the Jordan-Wigner 1D chain winds around.

We present an encoder ZX-diagram for each plaquette which enables directly reading off linearly independent logical and stabilizer Pauli operators that comprise the mapping\footnote{This observation is attributed to Ref.~\cite{huangZXstabgrok}. In systems exhibiting XY-, YZ-, or XYZ- symmetries, it is elegant to use the Y spider~\cite{Lang2012trichro,delafuente2024xyzruby}, but we opted not to here to reduce notational burden.}:
\begin{equation}\label{eq:sqAQMdecomp}
    \scalebox{0.9}{\input{LocalEncodings/sqAQMdecomp.tikz}}
\end{equation}
As both of the boxed Pauli operators mutually commute, their order in the composition is arbitrary\footnote{The dotted lines are necessary for the composition, but not inherent to the Pauli operators.}.
Here the $\pm$ labels are shorthand for $\pm \frac{\pi}{2}$.

\begin{remark}
    In the encoder perspective, in addition to the plaquette stabilizers provided in the original paper, we had to assign logical operators, i.e. Pauli strings acting on both logical qubits (i.e. post-Jordan-Wigner transform) and physical qubits (i.e. post-square lattice AQM). Any choice of linearly independent logical operators is valid so long as they commute with all stabilizers, and all possible such choices are equivalent up to a unitary transformation on the logical qubits.
\end{remark}

How this encoder acts on hopping terms matches those given in the original paper.
Relative to the Jordan-Wigner transform, the mapped hopping term operators acquire a correction term $\overline{\textsf{X}_1 \textsf{Z}_2} = \textsf{X}_1 \textsf{Z}_2 \textsf{Z}_{1'}$ and $\overline{\textsf{Z}_3 \textsf{X}_4} = \textsf{Z}_1 \textsf{X}_2 \textsf{Z}_{2'}$ (as well as replacing all X's with Y's here) at every plaquette.
This can be verified by pushing through the encoder or by direct calculation of the Pauli operators.

In summation, we assign linearly independent logical operators in such a way that they always act on qubits in the same plaquette.
Tiling this as in Figure~\ref{fig:sqAQM} gives a ZX-diagram for the square lattice AQM on lattices of any size.

%% file: sections/Conclusion.tex
In this paper, we conducted a detailed study of fermion-to-qubit mappings through the lens of the ZX-calculus.
We showed that the ZX-calculus provides a unified framework that bridges the disconnect between various approaches for these mappings.
Specifically, we analyzed the three main classes of mappings: linear encodings, ternary tree mappings, and local encodings.
By showing the correspondence of linear encodings and ternary tree mappings with the phase-free ZX diagrams, we proved that ternary trees always yield a linear encoding.
Finally, we showed how the different ways of describing local encodings become unified as a single ZX diagram.

A key direction for future work is to extend this graphical formalism to discover and classify new mappings.
Leveraging the strengths of ZX-calculus in optimization and compilation, we would like to develop algorithms for finding optimal mappings tailored for specific interaction graphs and qubit connectivity.
To that end, we propose leveraging ZX-based software tools such as \emph{PauliOpt}, which enables the search for optimal Clifford conjugation for Pauli polynomials\cite{gogioso2023annealing}.
Since these Clifford conjugations can define fermion-to-qubit mappings, repurposing such tools could offer a practical method for finding mappings optimized for particular Hamiltonians or simulation goals.

Another promising avenue is the interplay between fermion-to-qubit mappings and quantum error correcting codes.
As describing local encodings in the ZX-calculus uses the same methods as error correcting codes, we believe more results to be transferrable between the two using this shared language.
We hope to co-design mappings with error-correction schemes that are resource efficient for fault-tolerant quantum simulation.

Finally, we aim to generalize our approach beyond fermionic systems.
Recent extensions of the ZX-calculus to infinite-dimensional Hilbert spaces~\cite{shaikhFockedupZX2024} provide a foundation for developing an analogous framework for boson-to-qubit mappings~\cite{Somma2003qsimboson}.
A natural next step is to integrate these approaches within the mixed-dimensional ZX calculus~\cite{wangCompletenessQufiniteZXW2024,poorZXcalculusCompleteFiniteDimensional2024}, enabling the study of hybrid systems with fermion-boson interactions, such as the Hubbard-Holstein model used to study superconductivity.

%% file: sections/Proofs.tex
\subsection{Proofs of Section~\ref{sec:linear-encodings}, on Linear Encodings}\label{sec:proofs-linear-encodings}

\parityrecurrence*
\begin{proof}
    {\allowdisplaybreaks
    Define $P_k$ to be the matrix $E_k$ with its first column removed.
    \begin{align*}
    \input{ParityRecurrenceLHS.tikz} \quad & \overset{\hyperref[S1]{S1}}{=} \quad \input{ParityRecurrenceProof/ParityRecurrenceProof0.tikz}\\
    \overset{\hyperref[RCopy]{R Copy}}{=}\input{ParityRecurrenceProof/ParityRecurrenceProof1.tikz} \quad & \overset{\hyperref[S2]{S2}}{=} \quad \input{ParityRecurrenceProof/ParityRecurrenceProof2.tikz}\\
    \overset{\hyperref[GCopy]{GCopy}}{=}\input{ParityRecurrenceProof/ParityRecurrenceProof3.tikz} \quad & \overset{\hyperref[matmult]{matmult}}{=} \quad\input{ParityRecurrenceProof/ParityRecurrenceProof4.tikz}\\
    \overset{E_j def}{=}\input{ParityRecurrenceProof/ParityRecurrenceProof5.tikz} \quad & = \quad\input{ParityRecurrenceRHS.tikz}
    \end{align*}
    }
\end{proof}

\ParityOperatorShape*
\begin{proof}
    The encoding matrix $E_N$ of the parity encoding on $N$ qubits is the $N \times N$ matrix with ones on and below the diagonal and zeroes elsewhere.

    To compute an encoded $a_p^{Par}$, we work out that
    \begin{itemize}
        \item $E_Ne_p= \sum_{q\geq p}e_p$
        \item $e^T_p E_N^{-1}= e_{p-1}+e_p$ 
        \item $(E^{-1})^T \!1_{p-1}= e_{p-1}$
    \end{itemize}
    Where applicable, we define $e_{-1}$ to be the zero vector.
    Note that in the case $p=0$, $a^{Par}_0$ is given by 
    
    $$\input{ParityOperatorzero.tikz}$$
\end{proof}

\begin{lemma}\label{Jbkinv}
    Let $J$ be a row vector of all ones. Then $J\beta_k^{-1}=(0,\cdots,0,1)$.
\end{lemma}
    
    \begin{proof}    
    We prove this by induction.
    Base case:  $\beta_0^{-1}=(1)$, $J=(1)$, $J\beta_0^{-1}=1$.\\
    Note that entry $(J\beta_k^{-1})_i$ is the XOR of column $i$ of $\beta_k^{-1}$. 
    
    Assume the claim for $k$, and recall that
    
    $\beta_{k+1}^{-1}=\begin{bsmallmatrix} 
      \beta_{k}^{-1} & 0 \\ 
      B & \beta_{k}^{-1} 
    \end{bsmallmatrix}$
    
    Then $J_{k+1}\beta_{k+1}^{-1}=(0^{(1\times 2^k)},J_k) \begin{bsmallmatrix} 
      \beta_{k}^{-1} & 0 \\ 
      B & \beta_{k}^{-1} 
    \end{bsmallmatrix}= (J_k B,J_k\beta_{k}^{-1})=(0,\cdots,0,0,\cdots,0,1)$ as required.
    \end{proof}
  
\begin{lemma}\label{betakALemma}
    Let $\beta_k$ be the encoder for the Bravyi-Kitaev transform, and $A$ be the $(2^k \times 2^k)$ matrix with all ones on its last row and zeros elsewhere. Then 
    $$\input{betakA/betakALHS.tikz}=\input{betakA/betakARHS.tikz}$$
\end{lemma}

\begin{proof}
Let $N=2^k$, and note that $A=\begin{bsmallmatrix}
    \mathbf{0}^{(N-1)\times N}\\
    J^{1\times N}
\end{bsmallmatrix}$ where $J$ is the matrix of all ones.

\begin{align*}
    &\input{betakA/betakALHS.tikz}\quad \overset{inv}{=} \quad\input{betakA/betakAlemmaProof0.tikz}\quad \overset{S1}{=} \quad\input{betakA/betakAlemmaProof1.tikz}\quad \overset{GCopy}{=} \quad\input{betakA/betakAlemmaProof2.tikz} \\
    &\quad \overset{matmult}{=} \quad\input{betakA/betakAlemmaProof3.tikz}\quad \overset{0}{=} \quad\input{betakA/betakAlemmaProof35.tikz}\quad \overset{\ref{Jbkinv}}{=} \quad\input{betakA/betakAlemmaProof4.tikz} \\
    &\quad \overset{0}{=} \quad\input{betakA/betakAlemmaProof5.tikz} \quad \overset{fuse}{=} \quad \input{betakA/betakAlemmaProof6.tikz}
\end{align*}
\end{proof}

\bkrecurrence*
\begin{proof}
  \begin{align*}
    &\input{bkrecurrenceproof/bkplusoneLHS.tikz} \quad=\quad \input{bkrecurrenceproof/bkszxproof1.tikz} \quad = \quad \input{bkrecurrenceproof/bkszxproof2.tikz}\\
    &\quad \overset{\hyperref[0]{0}}{=} \quad\input{bkrecurrenceproof/bkszxproof3.tikz} \quad \overset{fuse}{=}\quad \input{bkrecurrenceproof/bkszxproof4.tikz} \quad \overset{\hyperref[GCopy]{GCopy}}{=}\quad \input{bkrecurrenceproof/bkszxproof5.tikz}\\ 
    &\quad \overset{\ref{betakALemma}}{=}\quad \input{bkrecurrenceproof/bkszxproof6.tikz} \quad \overset{Z,X}{=}\quad \input{bkrecurrenceproof/bkszxproof7.tikz} \quad \overset{fuse,OCM}{=}\quad \input{bkrecurrenceproof/bkszxproof8.tikz}
  \end{align*}
\end{proof}

\subsection{Proofs of Section~\ref{sec:elec}, on Electronic Hamiltonians} \label{sec:proofs-elec}

\begin{lemma}\label{lem:controlledparts}
Let $v\in \mathbb{F}^N_2$. Then the controlled diagrams of\\ 
\begin{center}
    \setlength\tabcolsep{1pt}
    \begin{tabular}{c c c c} 
        \input{HamiltonianTerms/pivgreen.tikz} &,  \input{HamiltonianTerms/pivred.tikz} &, \input{HamiltonianTerms/vprojpi.tikz} &, and \input{HamiltonianTerms/vproj.tikz} are \\
        \input{HamiltonianTerms/pivgreencontrolled.tikz} &, \input{HamiltonianTerms/pivredcontrolled.tikz} &, \input{HamiltonianTerms/vprojpicontrolled.tikz} &, and \input{HamiltonianTerms/vprojcontrolled.tikz} respectively
    \end{tabular}
\end{center}

\end{lemma}

\begin{proof}
    We prove the first and third statements. The second and fourth follow similarly.
    Let $k\in\{0,1\}$.
    
    We note that the upside down triangle 
    $$\input{ZXBackground/fliptriangledef.tikz}$$
    acts on the comptational basis as
    $$\input{triangledef0.tikz} \text{ and } \input{triangledef1.tikz}$$    

    Then 
    $$\input{HamiltonianTerms/pivgreenproof.tikz}$$
    Which proves the first statement.\\

    To prove the third statement, plug in $\ket{0}$ to get
    $$\input{HamiltonianTerms/vprojpiproof0.tikz}$$ 
    and $\ket{1}$ to get 
    $$\input{HamiltonianTerms/vprojpiproof1.tikz}$$ 
\end{proof}

\begin{lemma}\label{Edwin}
    Let $D$ be a linear map with controlled diagram $\tilde{D}$. Then 
    $$\input{Edwin.tikz}$$
\end{lemma}
\begin{proof}
    The proof of this lemma is given in \cite{AgnewCtrlZXW, Agnew_2023}.
\end{proof}

\begin{lemma}
    \input{ZXBackground/had0.tikz} = \input{ZXBackground/had1.tikz}
\end{lemma}

\begin{proposition}[The Number Operators]\label{prop:number}
The controlled diagram for $n_p$ is given by 
$$\input{NumberOperator/bkControlledNum.tikz}$$
\end{proposition}
  \begin{proof}
    Using $n_p=a_p^\dagger a_p$, and Proposition~\ref{prop:apEgoal} the diagram for $n_p$ is \\
    \begin{center}
        \setlength\tabcolsep{1pt}
        \begin{tabular}{ c c c }
            \input{NumberOperator/bkNumOperator.tikz} & $\qquad  \eqlabel{unfuse} \qquad $&\input{NumberOperator/bkNumOperator2.tikz} $\qquad \overset{\hyperref[GCopy]{GCopy}}{=}$ \\ 
            \input{NumberOperator/bkNumOperator3.tikz} & $\qquad  \overset{\hyperref[ZXWRules]{copy}}{=} \qquad$ &\input{NumberOperator/bkNumOperator4.tikz} $\qquad \overset{\hyperref[ZXWRules]{copy}}{=} $\\  
            \input{NumberOperator/bkNumOperator5.tikz} & $\qquad  \overset{\hyperref[Rpivect]{Rpivect}}{=} \qquad $&\input{NumberOperator/bkNumOperator6.tikz} $\qquad $  
        \end{tabular}
    \end{center}
    Where in the final equality, we have used that 

    $$v^X_p = E e_p  \text{   so   }  e_p E^{-1} v^X_p = e_p E^{-1} E e_p =1$$
Then the controlled diagram for $n_p$ follows from \ref{lem:controlledparts}.
\end{proof}


\begin{proposition}[The Exchange Operators]\label{prop:exchange}
    The controlled diagram for an exchange operator $a_i^\dagger a_j^\dagger a_j a_i$ is given by

    $$\input{ExchangeOperator/ControlledExchangeOp.tikz}$$

    Where we define 
    $$\scalebox{0.8}{\input{scalabletriangle.tikz}}$$
\end{proposition}
\begin{proof}
An exchange operator takes the form
$$a_i^\dagger a_j^\dagger a_j a_i$$
Applying the fermionic anti-commutation rules, first to $a_i$ and $a_j$, then to $a^\dagger_j$ and $a_i$ gives\\

$$a_i^\dagger a_j^\dagger a_j a_i=-a_i^\dagger a_j^\dagger a_i a_j= a_i^\dagger a_i a_j^\dagger a_j=n_in_j$$

Under an encoding $E$, we compute that 
$$n_in_j=\input{ExchangeOperator/ExchangeOp1.tikz}\quad =\quad\input{ExchangeOperator/ExchangeOp2.tikz}\quad=\quad\input{ExchangeOperator/ExchangeOp3.tikz}$$

We recall that the blank matrix arrow represents the matrix of all ones, so that 
$$\input{copymat2.tikz}$$
This map copies the computational basis $\{\ket{0}, \ket{1}\}$

Then, using the Lemma \ref{lem:controlledparts}, the controlled diagram for $n_in_j$ is given by 

$$\input{ExchangeOperator/ControlledExchangeOp.tikz}$$

\end{proof}

\begin{lemma}\label{lemma:aij}
The controlled diagram for $a_i^\dagger a_j$ is given by
$$\input{Excitation/controlledaij.tikz}$$
\end{lemma}

\begin{proof}
Without loss of generality, we  may take $i < j$, and compute $a_i^\dagger a_j$.
\begin{align*}
    &\input{Excitation/ExcitationOp1.tikz} & \overset{fuse}{=} &\quad\input{Excitation/ExcitationOp2.tikz}\\
    \overset{\hyperref[S1]{S1}}{=}\quad & \input{Excitation/ExcitationOp3.tikz} & \overset{copy}{=} &\quad\input{Excitation/ExcitationOp5.tikz}\\
    \overset{\hyperref[Rpivect]{Rpivect}}{=}\quad & \input{Excitation/ExcitationOp6.tikz} & = &\quad\input{Excitation/ExcitationOp7.tikz}
\end{align*}

Where we have used that $v^X_i= E e_i$, $v^X_j= E e_j$, and $v^Z_i=E^{-1^T}1_{i-1}$  so that when we pass the red $\pi$ through the matrix arrow,
$$\begin{bsmallmatrix} e_i^T \\  e_j^T \end{bsmallmatrix} \!\! E^{-1} v^X_j = \begin{bsmallmatrix} e_i^T \\  e_j^T \end{bsmallmatrix} \!\! E^{-1} E e_j = \begin{bsmallmatrix} 0 \\  1 \end{bsmallmatrix}$$

And we pick up no negative sign when swapping the green and red phases, as 
$$v^Z_i \cdot (v^X_i+v^X_j) = E^{-1^T} 1_{i-1} \cdot E(e_i+e_j)= 1_{i-1}^T E^{-1} E (e_i+e_j) = 0 \text{ as } j>i $$
Note that if instead $j<i$, we would pick up a $-1$ phase at this step

Applying \ref{lem:controlledparts}, The controlled diagram for $a_i^\dagger a_j$ is then 

$$\input{Excitation/controlledaij.tikz}$$
\end{proof}

\begin{proposition}[The Excitation Operators]\label{lem:excitation}
    The controlled diagram for the excitation operator $a_i^\dagger a_j+ a_j^\dagger a_i$ is given by 
    $$\scalebox{0.7}{\input{Excitation/controlledexc7.tikz}}$$
\end{proposition}
\begin{proof}
    We use the $W$ node to add the controlled diagrams of $a_i^\dagger a_j$ and its adjoint from Lemma~\ref{lemma:aij} , then apply ZX rewrite rules.
        \begin{align*}
        &\scalebox{0.7}{\input{Excitation/controlledexc.tikz}}\\
        &=\scalebox{0.7}{\input{Excitation/controlledexc2.tikz}}\\
        \overset{\ref{Edwin}}{=}&\scalebox{0.7}{\input{Excitation/controlledexc3.tikz}}
        =\scalebox{0.7}{\input{Excitation/controlledexc4.tikz}}\\
        \overset{\hyperref[matmult]{matmult}}{=}&\scalebox{0.7}{\input{Excitation/controlledexc5.tikz}}
        =\scalebox{0.7}{\input{Excitation/controlledexc6.tikz}}\\
        =&\scalebox{0.7}{\input{Excitation/controlledexc7.tikz}}
        \end{align*}
\end{proof}

\begin{proposition}[The Number-Excitation Operators]\label{prop:Number-Excitation}
The controlled diagram for $a_i^\dagger a_j^\dagger a_j a_k + a_k^\dagger a_j^\dagger a_j a_i$ is given by 
    $$\input{numberexop.tikz}$$
\end{proposition}
\begin{proof}
Using the fermionic anti-commutation relations, swapping the final two factors in each term, and then the middle two factors in each term gives

$$a_i^\dagger a_j^\dagger a_j a_k + a_k^\dagger a_j^\dagger a_j a_i= -a_i^\dagger a_j^\dagger a_k a_j - a_k^\dagger a_j^\dagger a_i a_j= a_i^\dagger a_k a_j^\dagger a_j + a_k^\dagger a_i a_j^\dagger a_j=(a_i^\dagger a_k + a_k^\dagger a_i )a_j^\dagger a_j$$

This is exactly the product of a number operator and an excitation operator, so we can immediately see that its controlled diagram is given by 

$$\scalebox{0.7}{\input{NumberExcitation/numexccontrolled.tikz}}$$
We can simplify this further, by applying the bialgebra rule in the bottom left of the diagram.

Note that for a row vector $v$, and a column vector $w$, applying the bialgebra rule gives us the following:
$$\scalebox{0.7}{\input{NumberExcitation/numexcbialg1.tikz}}=\scalebox{0.7}{\input{NumberExcitation/numexcbialg2.tikz}}=\scalebox{0.7}{\input{NumberExcitation/numexcbialg3.tikz}}=\scalebox{0.7}{\input{NumberExcitation/numexcbialg4.tikz}}$$
So in our case, $v=e_j^T E^{-1}$, and $w=E(e_i+e_k)$
so $v \cdot w =e_j^T E^{-1} E(e_i+e_k)=e_j^T (e_i+e_k)=0$ as $i,j,k$ are distinct. By the fact that the zero matrix arrow disconnects, we are left with 
$$\scalebox{0.7}{\input{NumberExcitation/numexcbialgfinal.tikz}}$$

Notice that the diagram on the left appears in the bottom right of the diagram for the number-excitation operator, so the lemma above allows us to swap the red and green spiders.
Then our number-excitation operator is given by
$$\scalebox{0.7}{\input{NumberExcitation/numexccontrolled2.tikz}} \quad = \quad \scalebox{0.7}{\input{NumberExcitation/numexccontrolled3.tikz}}$$
\end{proof}

\begin{proposition}[The Double Excitation Operators]\label{prop:Double-Excitation}
    The controlled diagram for the double excitation operator $a_i^\dagger a_j^\dagger a_k a_l +a_l^\dagger a_k^\dagger a_j a_i$ is
    \begin{equation*}
        \input{doubleexcop.tikz}
      \end{equation*}
\end{proposition}
\begin{proof}
    First, note that $a_i^\dagger a_j^\dagger a_k a_l=-a_i^\dagger a_j^\dagger a_l a_k=a_i^\dagger  a_l a_j^\dagger a_k$
Diagrammatically, this is given by 
$$\scalebox{0.7}{\input{DoubleExcitation/DoubleExcE1.tikz}}$$
Where $\{p<q\}$ is $0$ if $p<q$ and $1$ if $p>q$.\\ 
\begin{align*}
\scalebox{0.7}{\input{DoubleExcitation/DoubleExcE2.tikz}}
=\scalebox{0.7}{\input{DoubleExcitation/DoubleExcE3.tikz}}
\end{align*}
We introduce the shorthand
$v^Z_{ijkl}=v^Z_i+v^Z_j+ v^Z_k+v^Z_l$ and $v^X_{ijkl}=v^X_i+v^X_j+ v^X_k+v^X_l$

To write the term above as 

$$\scalebox{0.7}{\input{DoubleExcitation/DoubleExcE4.tikz}}$$

Note that the phase factor 
\begin{align*}
    (1_{i-1}+1_{l-1})\cdot (e_j+e_k) &=1_{i-1}\cdot e_j+1_{i-1}\cdot e_k + 1_{l-1}\cdot e_j + 1_{l-1}\cdot e_k\\
                                     &=\{i<j\}+\{i<k\}+\{l<j\} + \{l<k\}
\end{align*}
Hence, the phase factor in the double excitation term is the product over (unordered) pairs \{p,q\} of indices  $\prod_{\{p,q\}}(-1)^{\{p<q\}}$.

We implement this phase factor using a green spider labelled $b\pi$.

Using Lemma \ref{lem:controlledparts}, the controlled diagram for $a_i^\dagger a_j^\dagger a_k a_l$ is given by

$$\input{DoubleExcitation/DoubleExcHalf.tikz}$$

We use the $W$ to add the controlled diagram for $a_i^\dagger a_j^\dagger a_j a_k$ to its adjoint. This yields the following diagram for the double excitation operator, which we will simplify.

\begin{align*}
&\scalebox{0.7}{\input{DoubleExcitation/DoubleExcEsum.tikz}}
\overset{fuse}{=} \scalebox{0.7}{\input{DoubleExcitation/DoubleExcEsum2.tikz}}\\
\overset{\ref{Edwin}}{=} &\scalebox{0.7}{\input{DoubleExcitation/DoubleExcEsum3.tikz}}
\end{align*}

Next, we recall that $v^Z_{ijkl} = (1_{i-1}+1_{j-1}+1_{k-1}+1_{l-1})^T E^{-1}$, where we abbreviate $\mathbf{1}_{ijkl}:=1_{i-1}+1_{j-1}+1_{k-1}+1_{l-1}$.
{\allowdisplaybreaks
\begin{align*}
    = &\scalebox{0.7}{\input{DoubleExcitation/DoubleExcEsum4.tikz}}
    = \scalebox{0.7}{\input{DoubleExcitation/DoubleExcEsum5.tikz}}\\
    = &\scalebox{0.7}{\input{DoubleExcitation/DoubleExcEsum6.tikz}}
    =\scalebox{0.7}{\input{DoubleExcitation/DoubleExcEsum7.tikz}}
\end{align*}
}
\end{proof}

\subsection{Computing the Hamiltonian of the Hydrogen Molecule under the Bravyi-Kitaev encoding}
\label{HydrogenCoeffs}

The Hamiltonian of the hydrogen molecule takes the form
$$H=\sum_{ij} h_{ij} a^\dagger_ia_j+ \frac{1}{2} \sum_{ijkl} h_{ijkl} a^\dagger_ia^\dagger_j a_ka_l$$
where the coefficients $h_{ij}$ and $h_{ijkl}$ are given in the table below. All other coefficients are zero.
\begin{table}[h!]
    \begin{center}
    \begin{tabular}{ |c| c| }
    \hline
    Integrals & Value (atomic units) \\
    \hline
    $\mu_1:=h_{00} = h_{11}$ & $-1.252477$ \\
    \hline
    $\mu_2:=h_{22} = h_{33}$& $-0.475934$\\
    \hline
    $\mu_3:=h_{0110} = h_{1001}$ & $\ 0.674493$\\
    \hline
    $\mu_4:=h_{2332} = h_{3223}$ & $\ 0.697397$\\
    \hline
    $\quad \mu_5:=h_{0220} = h_{0330} = h_{1221} = h_{1331}$ & \multirow{2}{*}{$0.663472$}\\
    $= h_{2002} = h_{3003} = h_{2112} = h_{3113}$ & \\
    \hline
    $\mu_6:=h_{0202} = h_{1313} = h_{2130} = h_{2310} = h_{0312} = h_{0132}$ & $0.181287$\\
    \hline
    \end{tabular}
    \caption{The overlap integrals for molecular hydrogen in a minimal basis. This table is reprinted from Seeley et al.~\cite{Seeley_2012}.}
    \end{center}
\end{table}\\
Thus, given in full, the Hamiltonian of the Hydrogen molecule is
\begin{align}
    H \enspace = \enspace \quad & h_{00}a_0^\dagger a_0 +h_{11}a_1^\dagger a_1 +h_{22}a_2^\dagger a_2 +h_{33}a_3^\dagger a_3 \label{exp:1}\\
        +\; & h_{0110}a_0^\dagger a_1^\dagger a_1 a_0 + h_{2332}a_2^\dagger a_3^\dagger a_3 a_2+h_{0330}a_0^\dagger a_3^\dagger a_3 a_0 + h_{1221}a_1^\dagger a_2^\dagger a_2 a_1\label{exp:2}\\
        + \; &  (h_{0220}-h_{0202})a_0^\dagger a_2^\dagger a_2 a_0 \;+ \; (h_{1331}-h_{1313})a_1^\dagger a_3^\dagger a_3 a_1\label{exp:3}\\
        +\; &  h_{0132}(a_0^\dagger a_1^\dagger a_3 a_2 + a_2^\dagger a_3^\dagger a_1 a_0) \; + \; h_{0312}(a_0^\dagger a_3^\dagger a_1 a_2+a_2^\dagger a_1^\dagger a_3 a_0) \label{exp:4}
\end{align}

Under the Bravyi-Kitaev encoding, the fermionic annihilation operators take the form:\\
\begin{center}
    \begin{tabular}{ | c | c | c | c | c |}
    \hline
    $i$ & 0 & 1 & 2 & 3 \\
    \hline
    $a^{BK}_i$ & \input{HydrogenHamiltonian/a_0BK.tikz} & \input{HydrogenHamiltonian/a_1BK.tikz} & \input{HydrogenHamiltonian/a_2BK.tikz} & \input{HydrogenHamiltonian/a_3BK.tikz} \\
    \hline
    \end{tabular}
\end{center}
\subsubsection{Terms of $\ref{exp:1}$}

The first collection of terms is $\sum_{i} h_{ii} n_i$, where the $n_i$ are number operators.\\
The controlled diagram of  $h_{00}n_0+h_{11}n_1+h_{22}n_2+h_{33}n_3$ can be written as
$$\scalebox{0.8}{\input{HydrogenHamiltonian/H_0controlled1.tikz}} \qquad = \qquad \scalebox{0.8}{\input{HydrogenHamiltonian/H_0controlled2.tikz}}$$
\subsubsection{Terms of $\ref{exp:2}+\ref{exp:3}$}

Next, we compute the sum of the exchange operators in the Hamiltonian:\\
This is 
\begin{align*}
    & h_{0110}a_0^\dagger a_1^\dagger a_1 a_0 \enspace + \enspace h_{2332}a_2^\dagger a_3^\dagger a_3 a_2 \enspace + \enspace h_{0330}a_0^\dagger a_3^\dagger a_3 a_0 \enspace + \enspace h_{1221}a_1^\dagger a_2^\dagger a_2a_1\\ 
    + \; &(h_{0220}-h_{0202})a_0^\dagger a_2^\dagger a_2 a_0 \enspace + \enspace (h_{1331}-h_{1313})a_1^\dagger a_3^\dagger a_3 a_1
\end{align*}

Note that by the fermionic anti-commutation rules,
$$a_i^\dagger a_j^\dagger a_j a_i=-a_i^\dagger a_j^\dagger a_i a_j=a_i^\dagger  a_i a_j^\dagger a_i=n_in_j$$

rebracketing, we can express this sum as
\begin{align}
    &h_{0110}n_0n_1 + h_{2332}n_2n_3+h_{0330}n_0n_3 + h_{1221}n_1n_2 + (h_{0220}-h_{0202})n_0n_2  +  (h_{1331}-h_{1313})n_1n_3\\
    = &h_{0110}n_0n_1 + (h_{1221}n_1 + (h_{0220}-h_{0202})n_0)n_2+ (h_{2332}n_2+  (h_{1331}-h_{1313})n_1+h_{0330}n_0)n_3\\
    = &\mu_{3}n_0n_1 + (\mu_{5}n_1 + (\mu_5 - \mu_6)n_0)n_2+ (\mu_{4}n_2+  (\mu_5-\mu_6)n_1+\mu_5n_0)n_3
\end{align}
Note that this diagram uses three $W$ nodes, rather than the six we would've needed before rebracketing.
Here, the $W$ node signifies addition, while the $Z$ node signifies multiplication to derive the diagram
$$\scalebox{0.8}{\input{HydrogenHamiltonian/CoulombOperators.tikz}}$$
\subsubsection{Terms of $\ref{exp:4}$}

Finally, the double excitation terms are $h_{0132}(a_0^\dagger a_1^\dagger a_3 a_2+ a_2^\dagger a_3^\dagger a_1 a_0) \text{  and  } h_{0312}(a_0^\dagger a_3^\dagger a_1 a_2+ a_2^\dagger a_1^\dagger a_3 a_0)$.
$$D_1=a_0^\dagger a_1^\dagger a_3 a_2+ a_2^\dagger a_3^\dagger a_1 a_0 \text{ and } D_2=a_0^\dagger a_3^\dagger a_1 a_2+ a_2^\dagger a_1^\dagger a_3 a_0$$

Applying the result for double excitation operators gives us that the first of these terms is 
$$\scalebox{0.8}{\input{HydrogenHamiltonian/DoubleExc1controlled.tikz}}$$
We also note that under Bravyi-Kitaev, $D_2=-X_1D_1X_1$.
Applying ZXW rules gives us that the controlled diagram for $h_{0132}D_1+h_{0312}D_2$ is given by

$$\scalebox{0.8}{\input{HydrogenHamiltonian/DoubleExctotalcontrolled.tikz}}$$

Putting all these diagrams together gives us the hydrogen molecule under the Bravyi-Kitaev encoding.
$$\scalebox{0.7}{\input{HydrogenHamiltonian.tikz}}$$

\subsection{Proof in Section~\ref{sec:ternary-tree}}\label{sec:ternary-proofs}
\ternaryTreeCorrectness*
\begin{proof}
    The first case of induction is proved in the main text. Here we prove the remaining cases.
    \begin{enumerate}
        \setcounter{enumi}{1}
        \item Case: $k = i$\\
        For the majorana operator $\Gamma_{2k}$, we have
        \begin{equation*}
            \scalebox{0.8}{\input{TernaryTree/proof-root-x.tikz}}
        \end{equation*}
        For the majorana operator $\Gamma_{2k+1}$, we have
        \begin{equation*}
            \scalebox{0.8}{\input{TernaryTree/proof-root-y.tikz}}
        \end{equation*}
        \item Case: $k \gt a$ and $k \lt a+b+1$\\
        First we push the majorana operator $\Gamma_{2k+1}$. The proof for $\Gamma_{2k}$ is analogous.
        \begin{equation*}
            \scalebox{0.8}{\input{TernaryTree/proof-Y.tikz}}
        \end{equation*}
        where $k' = k - a$.
        \item Case: $k \gt a+b$\\
        First we push the majorana operator $\Gamma_{2k+1}$. The proof for $\Gamma_{2k}$ is analogous.
        \begin{equation*}
            \scalebox{0.8}{\input{TernaryTree/proof-Z.tikz}}
        \end{equation*}
        where $k' = k - (a+b)$.
    \end{enumerate}
\end{proof}

%% file: figures/HamiltonianTerms/pivgreen.tikz
\begin{tikzpicture}
	\begin{pgfonlayer}{nodelayer}
		\node [style=none] (0) at (-1, 0) {};
		\node [style=bold_gn] (1) at (0, 0) {};
		\node [style=none] (2) at (1, 0) {};
		\node [style=label] (3) at (0, 0.5) {$\pi v$};
	\end{pgfonlayer}
	\begin{pgfonlayer}{edgelayer}
		\draw [style=boldedge] (1) to (0.center);
		\draw [style=boldedge] (1) to (2.center);
	\end{pgfonlayer}
\end{tikzpicture}

%% file: figures/HamiltonianTerms/pivred.tikz
\begin{tikzpicture}
	\begin{pgfonlayer}{nodelayer}
		\node [style=none] (0) at (-1, 0) {};
		\node [style=bold_rn] (1) at (0, 0) {};
		\node [style=none] (2) at (1, 0) {};
		\node [style=label] (3) at (0, 0.5) {$\pi v$};
	\end{pgfonlayer}
	\begin{pgfonlayer}{edgelayer}
		\draw [style=boldedge] (1) to (0.center);
		\draw [style=boldedge] (1) to (2.center);
	\end{pgfonlayer}
\end{tikzpicture}

%% file: figures/ExchangeOperator/ControlledExchangeOp.tikz
\begin{tikzpicture}
	\begin{pgfonlayer}{nodelayer}
		\node [style=none] (0) at (-2.25, 0) {};
		\node [style={bold_gn}] (1) at (0, 0) {};
		\node [style=none] (2) at (2.25, 0) {};
		\node [style=label] (5) at (0.25, 2) {$2$};
		\node [style=label] (7) at (-1.5, 1) {$\begin{bsmallmatrix} e_i^T \\
  e_j^T 
\end{bsmallmatrix} \! E^{-1}$};
		\node [style=umat] (8) at (0, 1.25) {};
		\node [style=dtbold] (9) at (0, 2.75) {};
		\node [style=dmat] (10) at (0, 4) {};
		\node [style=none] (11) at (0, 5) {};
	\end{pgfonlayer}
	\begin{pgfonlayer}{edgelayer}
		\draw [style=boldedge] (1) to (0.center);
		\draw [style=boldedge] (2.center) to (1);
		\draw [style=boldedge] (1) to (8);
		\draw [style=boldedge] (9) to (10);
		\draw (10) to (11.center);
		\draw [style=boldedge] (8) to (9);
	\end{pgfonlayer}
\end{tikzpicture}

%% file: figures/ExchangeOperator/ExchangeOp1.tikz
\begin{tikzpicture}
	\begin{pgfonlayer}{nodelayer}
		\node [style=none] (0) at (-2.75, 0) {};
		\node [style=bold_gn] (1) at (-1, 0) {};
		\node [style=bold_gn] (2) at (1, 0) {};
		\node [style=none] (3) at (2.75, 0) {};
		\node [style={rn_phase}] (5) at (-1, -2.5) {$\pi$};
		\node [style={rn_phase}] (7) at (1, -2.5) {$\pi$};
		\node [style=label] (8) at (-2, -1.5) {$e^T_jE^{-1}$};
		\node [style=label] (9) at (2, -1.5) {$e^T_iE^{-1}$};
		\node [style=dmat] (10) at (-1, -1.5) {};
		\node [style=dmat] (11) at (1, -1.5) {};
	\end{pgfonlayer}
	\begin{pgfonlayer}{edgelayer}
		\draw [style=boldedge] (1) to (0.center);
		\draw [style=boldedge] (1) to (2);
		\draw [style=boldedge] (2) to (3.center);
		\draw [style=boldedge] (1) to (10);
		\draw [style=boldedge] (2) to (11);
		\draw (10) to (5);
		\draw (11) to (7);
	\end{pgfonlayer}
\end{tikzpicture}

%% file: figures/ExchangeOperator/ExchangeOp2.tikz
\begin{tikzpicture}
	\begin{pgfonlayer}{nodelayer}
		\node [style=none] (0) at (-2.75, 0) {};
		\node [style=bold_gn] (1) at (0, 0) {};
		\node [style=none] (3) at (2.75, 0) {};
		\node [style={rn_phase}] (12) at (-1, -3) {$\pi$};
		\node [style={rn_phase}] (13) at (1, -3) {$\pi$};
		\node [style=label] (14) at (-2, -2) {$e^T_jE^{-1}$};
		\node [style=label] (15) at (2, -2) {$e^T_iE^{-1}$};
		\node [style=dmat] (16) at (-1, -2) {};
		\node [style=dmat] (17) at (1, -2) {};
		\node [style=bold_gn] (18) at (0, -1) {};
	\end{pgfonlayer}
	\begin{pgfonlayer}{edgelayer}
		\draw [style=boldedge] (1) to (0.center);
		\draw [style=boldedge] (3.center) to (1);
		\draw (16) to (12);
		\draw (17) to (13);
		\draw [style=boldedge] (18) to (1);
		\draw [style=boldedge, bend right=45] (18) to (16);
		\draw [style=boldedge, bend left=45] (18) to (17);
	\end{pgfonlayer}
\end{tikzpicture}

%% file: figures/ExchangeOperator/ExchangeOp3.tikz
\begin{tikzpicture}
	\begin{pgfonlayer}{nodelayer}
		\node [style=none] (0) at (-2.75, 0) {};
		\node [style=bold_gn] (1) at (0, 0) {};
		\node [style=none] (2) at (2.75, 0) {};
		\node [style=rn] (3) at (0, -2.75) {};
		\node [style=dmat] (4) at (0, -1.5) {};
		\node [style=label] (5) at (-0.25, -2) {$2$};
		\node [style=label] (6) at (0.75, -2.75) {$ \pi \! \begin{bsmallmatrix} 1 \\
  1 
\end{bsmallmatrix}$};
		\node [style=label] (7) at (-1.25, -1.5) {$\begin{bsmallmatrix} e_i^T \\
  e_j^T 
\end{bsmallmatrix} \! E^{-1}$};
	\end{pgfonlayer}
	\begin{pgfonlayer}{edgelayer}
		\draw [style=boldedge] (1) to (0.center);
		\draw [style=boldedge] (1) to (4);
		\draw [style=boldedge] (4) to (3);
		\draw [style=boldedge] (2.center) to (1);
	\end{pgfonlayer}
\end{tikzpicture}

%% file: fermion-to-qubit.bbl
\begin{thebibliography}{10}
\providecommand{\bibitemdeclare}[2]{}
\providecommand{\surnamestart}{}
\providecommand{\surnameend}{}
\providecommand{\urlprefix}{Available at }
\providecommand{\url}[1]{\texttt{#1}}
\providecommand{\href}[2]{\texttt{#2}}
\providecommand{\urlalt}[2]{\href{#1}{#2}}
\providecommand{\doi}[1]{doi:\urlalt{https://doi.org/#1}{#1}}
\providecommand{\eprint}[1]{arXiv:\urlalt{https://arxiv.org/abs/#1}{#1}}
\providecommand{\bibinfo}[2]{#2}

\bibitemdeclare{mastersthesis}{Agnew_2023}
\bibitem{Agnew_2023}
\bibinfo{author}{Edwin \surnamestart Agnew\surnameend} (\bibinfo{year}{2023}):
  \emph{\bibinfo{title}{Quantum Polynomials in the ZXW Calculus}}.
\newblock Master's thesis, \bibinfo{school}{University of Oxford}.

\bibitemdeclare{article}{AgnewCtrlZXW}
\bibitem{AgnewCtrlZXW}
\bibinfo{author}{Edwin \surnamestart Agnew\surnameend}, \bibinfo{author}{Lia
  \surnamestart Yeh\surnameend} \& \bibinfo{author}{Richie \surnamestart
  Yeung\surnameend} (\bibinfo{year}{Manuscript in preparation}):
  \emph{\bibinfo{title}{Algebraic Structure of Controlled States and Operators
  in the ZXW-calculus}}.

\bibitemdeclare{article}{backensZXcalculusCompleteStabilizer2014}
\bibitem{backensZXcalculusCompleteStabilizer2014}
\bibinfo{author}{Miriam \surnamestart Backens\surnameend}
  (\bibinfo{year}{2014}): \emph{\bibinfo{title}{The {{ZX-calculus}} Is Complete
  for Stabilizer Quantum Mechanics}}.
\newblock {\slshape \bibinfo{journal}{New Journal of Physics}}
  \bibinfo{volume}{16}(\bibinfo{number}{9}), p. \bibinfo{pages}{093021},
  \doi{10.1088/1367-2630/16/9/093021}.

\bibitemdeclare{misc}{debeaudrapFastEffectiveTechniques2020}
\bibitem{debeaudrapFastEffectiveTechniques2020}
\bibinfo{author}{Niel \surnamestart de~Beaudrap\surnameend},
  \bibinfo{author}{Xiaoning \surnamestart Bian\surnameend} \&
  \bibinfo{author}{Quanlong \surnamestart Wang\surnameend}
  (\bibinfo{year}{2020}): \emph{\bibinfo{title}{Fast and effective techniques
  for T-count reduction via spider nest identities}}.
\newblock \eprint{2004.05164}.

\bibitemdeclare{inproceedings}{bonchiInteractingBialgebrasAre2014}
\bibitem{bonchiInteractingBialgebrasAre2014}
\bibinfo{author}{Filippo \surnamestart Bonchi\surnameend},
  \bibinfo{author}{Pawe{\l} \surnamestart Soboci{\'n}ski\surnameend} \&
  \bibinfo{author}{Fabio \surnamestart Zanasi\surnameend}
  (\bibinfo{year}{2014}): \emph{\bibinfo{title}{Interacting {{Bialgebras Are
  Frobenius}}}}.
\newblock In \bibinfo{editor}{Anca \surnamestart Muscholl\surnameend}, editor:
  {\slshape \bibinfo{booktitle}{Foundations of {{Software Science}} and
  {{Computation Structures}}}}, \bibinfo{series}{Lecture {{Notes}} in
  {{Computer Science}}}, \bibinfo{publisher}{Springer},
  \bibinfo{address}{Berlin, Heidelberg}, pp. \bibinfo{pages}{351--365},
  \doi{10.1007/978-3-642-54830-7_23}.

\bibitemdeclare{inproceedings}{borgnaEncodingHighlevelQuantum2023}
\bibitem{borgnaEncodingHighlevelQuantum2023}
\bibinfo{author}{Augustin \surnamestart Borgna\surnameend} \&
  \bibinfo{author}{Rafael \surnamestart Romero\surnameend}
  (\bibinfo{year}{2023}): \emph{\bibinfo{title}{Encoding High-Level Quantum
  Programs as {{SZX-diagrams}}}}.
\newblock In \bibinfo{editor}{Stefano \surnamestart Gogioso\surnameend} \&
  \bibinfo{editor}{Matty \surnamestart Hoban\surnameend}, editors: {\slshape
  \bibinfo{booktitle}{Proceedings 19th International Conference on Quantum
  Physics and Logic, Wolfson College, Oxford, {{UK}}, 27 June - 1 July 2022}},
  {\slshape \bibinfo{series}{Electronic Proceedings in Theoretical Computer
  Science}} \bibinfo{volume}{394}, \bibinfo{publisher}{Open Publishing
  Association}, pp. \bibinfo{pages}{141--169}, \doi{10.4204/EPTCS.394.9}.

\bibitemdeclare{article}{Bravyi_2002}
\bibitem{Bravyi_2002}
\bibinfo{author}{Sergey~B. \surnamestart Bravyi\surnameend} \&
  \bibinfo{author}{Alexei~Yu. \surnamestart Kitaev\surnameend}
  (\bibinfo{year}{2002}): \emph{\bibinfo{title}{Fermionic Quantum
  Computation}}.
\newblock {\slshape \bibinfo{journal}{Annals of Physics}}
  \bibinfo{volume}{298}(\bibinfo{number}{1}), p. \bibinfo{pages}{210–226},
  \doi{10.1006/aphy.2002.6254}.
\newblock \urlprefix\url{http://dx.doi.org/10.1006/aphy.2002.6254}.

\bibitemdeclare{inproceedings}{CaretteOracles2021}
\bibitem{CaretteOracles2021}
\bibinfo{author}{Titouan \surnamestart Carette\surnameend},
  \bibinfo{author}{Yohann \surnamestart D'Anello\surnameend} \&
  \bibinfo{author}{Simon \surnamestart Perdrix\surnameend}
  (\bibinfo{year}{2021}): \emph{\bibinfo{title}{{Quantum Algorithms and Oracles
  with the Scalable ZX-calculus}}}.
\newblock In \bibinfo{editor}{Chris \surnamestart Heunen\surnameend} \&
  \bibinfo{editor}{Miriam \surnamestart Backens\surnameend}, editors: {\slshape
  \bibinfo{booktitle}{Proceedings 18th International Conference on Quantum
  Physics and Logic, Gdansk, Poland, and online, 7-11 June 2021}}, {\slshape
  \bibinfo{series}{Electronic Proceedings in Theoretical Computer Science}}
  \bibinfo{volume}{343}, \bibinfo{publisher}{Open Publishing Association}, pp.
  \bibinfo{pages}{193--209}, \doi{10.4204/EPTCS.343.10}.

\bibitemdeclare{inproceedings}{caretteSZXCalculusScalableGraphical2019}
\bibitem{caretteSZXCalculusScalableGraphical2019}
\bibinfo{author}{Titouan \surnamestart Carette\surnameend},
  \bibinfo{author}{Dominic \surnamestart Horsman\surnameend} \&
  \bibinfo{author}{Simon \surnamestart Perdrix\surnameend}
  (\bibinfo{year}{2019}): \emph{\bibinfo{title}{{{SZX-Calculus}}: {{Scalable
  Graphical Quantum Reasoning}}}}.
\newblock In \bibinfo{editor}{Peter \surnamestart Rossmanith\surnameend},
  \bibinfo{editor}{Pinar \surnamestart Heggernes\surnameend} \&
  \bibinfo{editor}{Joost-Pieter \surnamestart Katoen\surnameend}, editors:
  {\slshape \bibinfo{booktitle}{44th {{International Symposium}} on
  {{Mathematical Foundations}} of {{Computer Science}} ({{MFCS}} 2019)}},
  {\slshape \bibinfo{series}{Leibniz {{International Proceedings}} in
  {{Informatics}} ({{LIPIcs}})}} \bibinfo{volume}{138},
  \bibinfo{publisher}{Schloss Dagstuhl--Leibniz-Zentrum fuer Informatik},
  \bibinfo{address}{Dagstuhl, Germany}, pp. \bibinfo{pages}{55:1--55:15},
  \doi{10.4230/LIPIcs.MFCS.2019.55}.

\bibitemdeclare{article}{ChenXu2023fermiontoqubit}
\bibitem{ChenXu2023fermiontoqubit}
\bibinfo{author}{Yu-An \surnamestart Chen\surnameend} \& \bibinfo{author}{Yijia
  \surnamestart Xu\surnameend} (\bibinfo{year}{2023}):
  \emph{\bibinfo{title}{Equivalence between Fermion-to-Qubit Mappings in two
  Spatial Dimensions}}.
\newblock {\slshape \bibinfo{journal}{PRX Quantum}} \bibinfo{volume}{4}, p.
  \bibinfo{pages}{010326}, \doi{10.1103/PRXQuantum.4.010326}.
\newblock \urlprefix\url{https://link.aps.org/doi/10.1103/PRXQuantum.4.010326}.

\bibitemdeclare{misc}{chiewTernaryTreeTransformations2024a}
\bibitem{chiewTernaryTreeTransformations2024a}
\bibinfo{author}{Mitchell \surnamestart Chiew\surnameend},
  \bibinfo{author}{Brent \surnamestart Harrison\surnameend} \&
  \bibinfo{author}{Sergii \surnamestart Strelchuk\surnameend}
  (\bibinfo{year}{2024}): \emph{\bibinfo{title}{Ternary Tree Transformations
  Are Equivalent to Linear Encodings of the {{Fock}} Basis}},
  \doi{10.48550/arXiv.2412.07578}.
\newblock \eprint{2412.07578}.

\bibitemdeclare{article}{chiew2023JWoptimal}
\bibitem{chiew2023JWoptimal}
\bibinfo{author}{Mitchell \surnamestart Chiew\surnameend} \&
  \bibinfo{author}{Sergii \surnamestart Strelchuk\surnameend}
  (\bibinfo{year}{2023}): \emph{\bibinfo{title}{Discovering optimal
  fermion-qubit mappings through algorithmic enumeration}}.
\newblock {\slshape \bibinfo{journal}{Quantum}} \bibinfo{volume}{7}, p.
  \bibinfo{pages}{1145}, \doi{10.22331/q-2023-10-18-1145}.
\newblock \urlprefix\url{http://dx.doi.org/10.22331/q-2023-10-18-1145}.

\bibitemdeclare{inproceedings}{coeckeInteractingQuantumObservables2008}
\bibitem{coeckeInteractingQuantumObservables2008}
\bibinfo{author}{Bob \surnamestart Coecke\surnameend} \& \bibinfo{author}{Ross
  \surnamestart Duncan\surnameend} (\bibinfo{year}{2008}):
  \emph{\bibinfo{title}{Interacting {{Quantum Observables}}}}.
\newblock In \bibinfo{editor}{Luca \surnamestart Aceto\surnameend},
  \bibinfo{editor}{Ivan \surnamestart Damg{\aa}rd\surnameend},
  \bibinfo{editor}{Leslie~Ann \surnamestart Goldberg\surnameend},
  \bibinfo{editor}{Magn{\'u}s~M. \surnamestart Halld{\'o}rsson\surnameend},
  \bibinfo{editor}{Anna \surnamestart Ing{\'o}lfsd{\'o}ttir\surnameend} \&
  \bibinfo{editor}{Igor \surnamestart Walukiewicz\surnameend}, editors:
  {\slshape \bibinfo{booktitle}{Automata, {{Languages}} and {{Programming}}}},
  \bibinfo{series}{Lecture {{Notes}} in {{Computer Science}}},
  \bibinfo{publisher}{Springer}, \bibinfo{address}{Berlin, Heidelberg}, pp.
  \bibinfo{pages}{298--310}, \doi{10.1007/978-3-540-70583-3_25}.
\newblock
  \urlprefix\url{http://personal.strath.ac.uk/ross.duncan/papers/iqo-icalp.pdf}.

\bibitemdeclare{article}{coeckeThreeQubitEntanglement2011}
\bibitem{coeckeThreeQubitEntanglement2011}
\bibinfo{author}{Bob \surnamestart Coecke\surnameend} \& \bibinfo{author}{Bill
  \surnamestart Edwards\surnameend} (\bibinfo{year}{2011}):
  \emph{\bibinfo{title}{Three Qubit Entanglement within Graphical
  {{Z}}/{{X-calculus}}}}.
\newblock {\slshape \bibinfo{journal}{Electronic Proceedings in Theoretical
  Computer Science}} \bibinfo{volume}{52}, pp. \bibinfo{pages}{22--33},
  \doi{10.4204/EPTCS.52.3}.

\bibitemdeclare{article}{debeaudrapZXCalculusLanguage2020}
\bibitem{debeaudrapZXCalculusLanguage2020}
\bibinfo{author}{Niel \surnamestart {de Beaudrap}\surnameend} \&
  \bibinfo{author}{Dominic \surnamestart Horsman\surnameend}
  (\bibinfo{year}{2020}): \emph{\bibinfo{title}{The {{ZX}} Calculus Is a
  Language for Surface Code Lattice Surgery}}.
\newblock {\slshape \bibinfo{journal}{Quantum}} \bibinfo{volume}{4}, p.
  \bibinfo{pages}{218}, \doi{10.22331/q-2020-01-09-218}.

\bibitemdeclare{inproceedings}{defeliceLightMatterInteractionZXW2023}
\bibitem{defeliceLightMatterInteractionZXW2023}
\bibinfo{author}{Giovanni \surnamestart {de Felice}\surnameend},
  \bibinfo{author}{Razin~A. \surnamestart Shaikh\surnameend},
  \bibinfo{author}{Boldizs{\'a}r \surnamestart Po{\'o}r\surnameend},
  \bibinfo{author}{Lia \surnamestart Yeh\surnameend}, \bibinfo{author}{Quanlong
  \surnamestart Wang\surnameend} \& \bibinfo{author}{Bob \surnamestart
  Coecke\surnameend} (\bibinfo{year}{2023}):
  \emph{\bibinfo{title}{Light-{{Matter Interaction}} in the {{ZXW Calculus}}}}.
\newblock In \bibinfo{editor}{Shane \surnamestart Mansfield\surnameend},
  \bibinfo{editor}{Benoit \surnamestart Val{\^i}ron\surnameend} \&
  \bibinfo{editor}{Vladimir \surnamestart Zamdzhiev\surnameend}, editors:
  {\slshape \bibinfo{booktitle}{Proceedings of the Twentieth International
  Conference on Quantum Physics and Logic, Paris, France, 17-21st July 2023}},
  {\slshape \bibinfo{series}{Electronic Proceedings in Theoretical Computer
  Science}} \bibinfo{volume}{384}, \bibinfo{publisher}{Open Publishing
  Association}, pp. \bibinfo{pages}{20--46}, \doi{10.4204/EPTCS.384.2}.

\bibitemdeclare{article}{Derby2021compact}
\bibitem{Derby2021compact}
\bibinfo{author}{Charles \surnamestart Derby\surnameend}, \bibinfo{author}{Joel
  \surnamestart Klassen\surnameend}, \bibinfo{author}{Johannes \surnamestart
  Bausch\surnameend} \& \bibinfo{author}{Toby \surnamestart Cubitt\surnameend}
  (\bibinfo{year}{2021}): \emph{\bibinfo{title}{Compact fermion to qubit
  mappings}}.
\newblock {\slshape \bibinfo{journal}{Physical Review B}}
  \bibinfo{volume}{104}(\bibinfo{number}{3}),
  \doi{10.1103/physrevb.104.035118}.
\newblock \urlprefix\url{http://dx.doi.org/10.1103/PhysRevB.104.035118}.

\bibitemdeclare{inproceedings}{dundar-coeckeQuantumPicturalismLearning2023}
\bibitem{dundar-coeckeQuantumPicturalismLearning2023}
\bibinfo{author}{Selma \surnamestart {D{\"u}ndar-Coecke}\surnameend},
  \bibinfo{author}{Lia \surnamestart Yeh\surnameend}, \bibinfo{author}{Caterina
  \surnamestart Puca\surnameend}, \bibinfo{author}{Sieglinde M.-L.
  \surnamestart Pfaendler\surnameend}, \bibinfo{author}{Muhammad~Hamza
  \surnamestart Waseem\surnameend}, \bibinfo{author}{Thomas \surnamestart
  Cervoni\surnameend}, \bibinfo{author}{Aleks \surnamestart
  Kissinger\surnameend}, \bibinfo{author}{Stefano \surnamestart
  Gogioso\surnameend} \& \bibinfo{author}{Bob \surnamestart Coecke\surnameend}
  (\bibinfo{year}{2023}): \emph{\bibinfo{title}{Quantum {{Picturalism}}:
  {{Learning Quantum Theory}} in {{High School}}}}.
\newblock In: {\slshape \bibinfo{booktitle}{2023 {{IEEE International
  Conference}} on {{Quantum Computing}} and {{Engineering}} ({{QCE}})}},
  \bibinfo{volume}{03}, pp. \bibinfo{pages}{21--32},
  \doi{10.1109/QCE57702.2023.20321}.
\newblock \eprint{2312.03653}.

\bibitemdeclare{article}{eastAKLTStatesZXDiagramsDiagrammatic2022}
\bibitem{eastAKLTStatesZXDiagramsDiagrammatic2022}
\bibinfo{author}{Richard~D.P. \surnamestart East\surnameend},
  \bibinfo{author}{John \surnamestart {van de Wetering}\surnameend},
  \bibinfo{author}{Nicholas \surnamestart Chancellor\surnameend} \&
  \bibinfo{author}{Adolfo~G. \surnamestart Grushin\surnameend}
  (\bibinfo{year}{2022}): \emph{\bibinfo{title}{{{AKLT-States}} as
  {{ZX-Diagrams}}: {{Diagrammatic Reasoning}} for {{Quantum States}}}}.
\newblock {\slshape \bibinfo{journal}{PRX Quantum}}
  \bibinfo{volume}{3}(\bibinfo{number}{1}), p. \bibinfo{pages}{010302},
  \doi{10.1103/PRXQuantum.3.010302}.

\bibitemdeclare{misc}{delafuente2024xyzruby}
\bibitem{delafuente2024xyzruby}
\bibinfo{author}{Julio C.~Magdalena \surnamestart de~la Fuente\surnameend},
  \bibinfo{author}{Josias \surnamestart Old\surnameend}, \bibinfo{author}{Alex
  \surnamestart Townsend-Teague\surnameend}, \bibinfo{author}{Manuel
  \surnamestart Rispler\surnameend}, \bibinfo{author}{Jens \surnamestart
  Eisert\surnameend} \& \bibinfo{author}{Markus \surnamestart
  Müller\surnameend} (\bibinfo{year}{2024}): \emph{\bibinfo{title}{The XYZ
  ruby code: Making a case for a three-colored graphical calculus for quantum
  error correction in spacetime}}.
\newblock \eprint{2407.08566}.

\bibitemdeclare{article}{gogioso2023annealing}
\bibitem{gogioso2023annealing}
\bibinfo{author}{Stefano \surnamestart Gogioso\surnameend} \&
  \bibinfo{author}{Richie \surnamestart Yeung\surnameend}
  (\bibinfo{year}{2023}): \emph{\bibinfo{title}{Annealing Optimisation of Mixed
  ZX Phase Circuits}}.
\newblock {\slshape \bibinfo{journal}{Electronic Proceedings in Theoretical
  Computer Science}} \bibinfo{volume}{394}, pp. \bibinfo{pages}{415--431},
  \doi{10.4204/eptcs.394.20}.

\bibitemdeclare{misc}{gorantla2024tensornetworksnoninvertiblesymmetries}
\bibitem{gorantla2024tensornetworksnoninvertiblesymmetries}
\bibinfo{author}{Pranay \surnamestart Gorantla\surnameend},
  \bibinfo{author}{Shu-Heng \surnamestart Shao\surnameend} \&
  \bibinfo{author}{Nathanan \surnamestart Tantivasadakarn\surnameend}
  (\bibinfo{year}{2024}): \emph{\bibinfo{title}{Tensor networks for
  non-invertible symmetries in 3+1d and beyond}}.
\newblock \eprint{2406.12978}.

\bibitemdeclare{misc}{harrison2024sierpinskitrianglefermiontoqubittransform}
\bibitem{harrison2024sierpinskitrianglefermiontoqubittransform}
\bibinfo{author}{Brent \surnamestart Harrison\surnameend},
  \bibinfo{author}{Mitchell \surnamestart Chiew\surnameend},
  \bibinfo{author}{Jason \surnamestart Necaise\surnameend},
  \bibinfo{author}{Andrew \surnamestart Projansky\surnameend},
  \bibinfo{author}{Sergii \surnamestart Strelchuk\surnameend} \&
  \bibinfo{author}{James~D. \surnamestart Whitfield\surnameend}
  (\bibinfo{year}{2024}): \emph{\bibinfo{title}{A Sierpinski Triangle
  Fermion-to-Qubit Transform}}.
\newblock \eprint{2409.04348}.

\bibitemdeclare{article}{huangZXstabgrok}
\bibitem{huangZXstabgrok}
\bibinfo{author}{Jiaxin \surnamestart Huang\surnameend}, \bibinfo{author}{Aleks
  \surnamestart Kissinger\surnameend}, \bibinfo{author}{Sarah~Meng
  \surnamestart Li\surnameend}, \bibinfo{author}{John \surnamestart van~de
  Wetering\surnameend} \& \bibinfo{author}{Lia \surnamestart Yeh\surnameend}
  (\bibinfo{year}{Manuscript in preparation}): \emph{\bibinfo{title}{ZX Normal
  Forms for Stabilizer Codes (…or How to Graphically Grok Tableaus)}}.

\bibitemdeclare{inproceedings}{huangGraphicalCSSCode2023}
\bibitem{huangGraphicalCSSCode2023}
\bibinfo{author}{Jiaxin \surnamestart Huang\surnameend},
  \bibinfo{author}{Sarah~Meng \surnamestart Li\surnameend},
  \bibinfo{author}{Lia \surnamestart Yeh\surnameend}, \bibinfo{author}{Aleks
  \surnamestart Kissinger\surnameend}, \bibinfo{author}{Michele \surnamestart
  Mosca\surnameend} \& \bibinfo{author}{Michael \surnamestart
  Vasmer\surnameend} (\bibinfo{year}{2023}): \emph{\bibinfo{title}{Graphical
  {{CSS}} Code Transformation Using {{ZX}} Calculus}}.
\newblock In \bibinfo{editor}{Shane \surnamestart Mansfield\surnameend},
  \bibinfo{editor}{Benoit \surnamestart Val{\^i}ron\surnameend} \&
  \bibinfo{editor}{Vladimir \surnamestart Zamdzhiev\surnameend}, editors:
  {\slshape \bibinfo{booktitle}{Proceedings of the Twentieth International
  Conference on Quantum Physics and Logic}}, {\slshape
  \bibinfo{series}{Electronic Proceedings in Theoretical Computer Science}}
  \bibinfo{volume}{384}, \bibinfo{publisher}{Open Publishing Association}, pp.
  \bibinfo{pages}{1--19}, \doi{10.4204/EPTCS.384.1}.

\bibitemdeclare{article}{jiangOptimalFermiontoqubitMapping2020}
\bibitem{jiangOptimalFermiontoqubitMapping2020}
\bibinfo{author}{Zhang \surnamestart Jiang\surnameend}, \bibinfo{author}{Amir
  \surnamestart Kalev\surnameend}, \bibinfo{author}{Wojciech \surnamestart
  Mruczkiewicz\surnameend} \& \bibinfo{author}{Hartmut \surnamestart
  Neven\surnameend} (\bibinfo{year}{2020}): \emph{\bibinfo{title}{Optimal
  Fermion-to-Qubit Mapping via Ternary Trees with Applications to Reduced
  Quantum States Learning}}.
\newblock {\slshape \bibinfo{journal}{Quantum}} \bibinfo{volume}{4}, p.
  \bibinfo{pages}{276}, \doi{10.22331/q-2020-06-04-276}.

\bibitemdeclare{article}{jordanUeberPaulischeAequivalenzverbot1928}
\bibitem{jordanUeberPaulischeAequivalenzverbot1928}
\bibinfo{author}{P.~\surnamestart Jordan\surnameend} \&
  \bibinfo{author}{E.~\surnamestart Wigner\surnameend} (\bibinfo{year}{1928}):
  \emph{\bibinfo{title}{{{\"U}ber das Paulische {\"A}quivalenzverbot}}}.
\newblock {\slshape \bibinfo{journal}{Zeitschrift f{\"u}r Physik}}
  \bibinfo{volume}{47}(\bibinfo{number}{9}), pp. \bibinfo{pages}{631--651},
  \doi{10.1007/BF01331938}.

\bibitemdeclare{misc}{kissingerPhasefreeZXDiagrams2022}
\bibitem{kissingerPhasefreeZXDiagrams2022}
\bibinfo{author}{Aleks \surnamestart Kissinger\surnameend}
  (\bibinfo{year}{2022}): \emph{\bibinfo{title}{Phase-free ZX diagrams are CSS
  codes (...or how to graphically grok the surface code)}}.
\newblock \eprint{2204.14038}.

\bibitemdeclare{article}{kissingerReducingTcountZXcalculus2020}
\bibitem{kissingerReducingTcountZXcalculus2020}
\bibinfo{author}{Aleks \surnamestart Kissinger\surnameend} \&
  \bibinfo{author}{John \surnamestart {van de Wetering}\surnameend}
  (\bibinfo{year}{2020}): \emph{\bibinfo{title}{Reducing {{T-count}} with the
  {{ZX-calculus}}}}.
\newblock {\slshape \bibinfo{journal}{Physical Review A}}
  \bibinfo{volume}{102}(\bibinfo{number}{2}), p. \bibinfo{pages}{022406},
  \doi{10.1103/PhysRevA.102.022406}.
\newblock \eprint{1903.10477}.

\bibitemdeclare{misc}{kissingerClassicalSimulationQuantum2022}
\bibitem{kissingerClassicalSimulationQuantum2022}
\bibinfo{author}{Aleks \surnamestart Kissinger\surnameend},
  \bibinfo{author}{John \surnamestart {van de Wetering}\surnameend} \&
  \bibinfo{author}{Renaud \surnamestart Vilmart\surnameend}
  (\bibinfo{year}{2022}): \emph{\bibinfo{title}{Classical Simulation of Quantum
  Circuits with Partial and Graphical Stabiliser Decompositions}},
  \doi{10.4230/LIPIcs.TQC.2022.5}.

\bibitemdeclare{article}{kissingerUniversalMBQCGeneralised2019}
\bibitem{kissingerUniversalMBQCGeneralised2019}
\bibinfo{author}{Aleks \surnamestart Kissinger\surnameend} \&
  \bibinfo{author}{John \surnamestart van~de Wetering\surnameend}
  (\bibinfo{year}{2019}): \emph{\bibinfo{title}{Universal {{MBQC}} with
  Generalised Parity-Phase Interactions and {{Pauli}} Measurements}}.
\newblock {\slshape \bibinfo{journal}{Quantum}} \bibinfo{volume}{3}, p.
  \bibinfo{pages}{134}, \doi{10.22331/q-2019-04-26-134}.

\bibitemdeclare{book}{KissingerWetering2024Book}
\bibitem{KissingerWetering2024Book}
\bibinfo{author}{Aleks \surnamestart Kissinger\surnameend} \&
  \bibinfo{author}{John \surnamestart van~de Wetering\surnameend}
  (\bibinfo{year}{2024}): \emph{\bibinfo{title}{{Picturing Quantum Software: An
  Introduction to the ZX-Calculus and Quantum Compilation}}}.
\newblock \bibinfo{publisher}{Preprint}.

\bibitemdeclare{article}{KissingerScalable2024}
\bibitem{KissingerScalable2024}
\bibinfo{author}{Aleks \surnamestart Kissinger\surnameend} \&
  \bibinfo{author}{John \surnamestart van~de Wetering\surnameend}
  (\bibinfo{year}{2024}): \emph{\bibinfo{title}{Scalable Spider Nests (...Or
  How to Graphically Grok Transversal Non-Clifford Gates)}}.
\newblock {\slshape \bibinfo{journal}{Electronic Proceedings in Theoretical
  Computer Science}} \bibinfo{volume}{406}, pp. \bibinfo{pages}{79--95},
  \doi{10.4204/eptcs.406.4}.

\bibitemdeclare{article}{Lang2012trichro}
\bibitem{Lang2012trichro}
\bibinfo{author}{Alex \surnamestart Lang\surnameend} \& \bibinfo{author}{Bob
  \surnamestart Coecke\surnameend} (\bibinfo{year}{2012}):
  \emph{\bibinfo{title}{Trichromatic Open Digraphs for Understanding Qubits}}.
\newblock {\slshape \bibinfo{journal}{Electronic Proceedings in Theoretical
  Computer Science}} \bibinfo{volume}{95}, p. \bibinfo{pages}{193–209},
  \doi{10.4204/eptcs.95.14}.
\newblock \urlprefix\url{http://dx.doi.org/10.4204/EPTCS.95.14}.

\bibitemdeclare{inproceedings}{mcelvanneyFlowpreservingZXcalculusRewrite2023}
\bibitem{mcelvanneyFlowpreservingZXcalculusRewrite2023}
\bibinfo{author}{Tommy \surnamestart McElvanney\surnameend} \&
  \bibinfo{author}{Miriam \surnamestart Backens\surnameend}
  (\bibinfo{year}{2023}): \emph{\bibinfo{title}{Flow-Preserving {{ZX-calculus}}
  Rewrite Rules for Optimisation and Obfuscation}}.
\newblock In \bibinfo{editor}{Shane \surnamestart Mansfield\surnameend},
  \bibinfo{editor}{Benoit \surnamestart Val{\^i}ron\surnameend} \&
  \bibinfo{editor}{Vladimir \surnamestart Zamdzhiev\surnameend}, editors:
  {\slshape \bibinfo{booktitle}{Proceedings of the Twentieth International
  Conference on Quantum Physics and Logic, Paris, France, 17-21st July 2023}},
  {\slshape \bibinfo{series}{Electronic Proceedings in Theoretical Computer
  Science}} \bibinfo{volume}{384}, \bibinfo{publisher}{Open Publishing
  Association}, pp. \bibinfo{pages}{203--219}, \doi{10.4204/EPTCS.384.12}.

\bibitemdeclare{misc}{millerTreespilationArchitectureStateOptimised2024}
\bibitem{millerTreespilationArchitectureStateOptimised2024}
\bibinfo{author}{Aaron \surnamestart Miller\surnameend}, \bibinfo{author}{Adam
  \surnamestart Glos\surnameend} \& \bibinfo{author}{Zolt{\'a}n \surnamestart
  Zimbor{\'a}s\surnameend} (\bibinfo{year}{2024}):
  \emph{\bibinfo{title}{Treespilation: {{Architecture-}} and {{State-Optimised
  Fermion-to-Qubit Mappings}}}}, \doi{10.48550/arXiv.2403.03992}.
\newblock \eprint{2403.03992}.

\bibitemdeclare{article}{millerBonsaiAlgorithmGrow2023}
\bibitem{millerBonsaiAlgorithmGrow2023}
\bibinfo{author}{Aaron \surnamestart Miller\surnameend},
  \bibinfo{author}{Zolt{\'a}n \surnamestart Zimbor{\'a}s\surnameend},
  \bibinfo{author}{Stefan \surnamestart Knecht\surnameend},
  \bibinfo{author}{Sabrina \surnamestart Maniscalco\surnameend} \&
  \bibinfo{author}{Guillermo \surnamestart {Garc{\'i}a-P{\'e}rez}\surnameend}
  (\bibinfo{year}{2023}): \emph{\bibinfo{title}{Bonsai {{Algorithm}}: {{Grow
  Your Own Fermion-to-Qubit Mappings}}}}.
\newblock {\slshape \bibinfo{journal}{PRX Quantum}}
  \bibinfo{volume}{4}(\bibinfo{number}{3}), p. \bibinfo{pages}{030314},
  \doi{10.1103/PRXQuantum.4.030314}.

\bibitemdeclare{article}{ngDiagrammaticCalculusFermionic2019}
\bibitem{ngDiagrammaticCalculusFermionic2019}
\bibinfo{author}{Kang~Feng \surnamestart Ng\surnameend}, \bibinfo{author}{Amar
  \surnamestart Hadzihasanovic\surnameend} \& \bibinfo{author}{Giovanni
  \surnamestart {de Felice}\surnameend} (\bibinfo{year}{2019}):
  \emph{\bibinfo{title}{A Diagrammatic Calculus of Fermionic Quantum
  Circuits}}.
\newblock {\slshape \bibinfo{journal}{Logical Methods in Computer Science}}
  \bibinfo{volume}{Volume 15, Issue 3}, \doi{10.23638/LMCS-15(3:26)2019}.

\bibitemdeclare{misc}{nigmatullin2024experimentalcompact}
\bibitem{nigmatullin2024experimentalcompact}
\bibinfo{author}{Ramil \surnamestart Nigmatullin\surnameend},
  \bibinfo{author}{Kevin \surnamestart Hemery\surnameend},
  \bibinfo{author}{Khaldoon \surnamestart Ghanem\surnameend},
  \bibinfo{author}{Steven \surnamestart Moses\surnameend}, \bibinfo{author}{Dan
  \surnamestart Gresh\surnameend}, \bibinfo{author}{Peter \surnamestart
  Siegfried\surnameend}, \bibinfo{author}{Michael \surnamestart
  Mills\surnameend}, \bibinfo{author}{Thomas \surnamestart
  Gatterman\surnameend}, \bibinfo{author}{Nathan \surnamestart
  Hewitt\surnameend}, \bibinfo{author}{Etienne \surnamestart Granet\surnameend}
  \& \bibinfo{author}{Henrik \surnamestart Dreyer\surnameend}
  (\bibinfo{year}{2024}): \emph{\bibinfo{title}{Experimental Demonstration of
  Break-Even for the Compact Fermionic Encoding}}.
\newblock \eprint{2409.06789}.

\bibitemdeclare{article}{OBrien2024ultrafast}
\bibitem{OBrien2024ultrafast}
\bibinfo{author}{Oliver \surnamestart O'Brien\surnameend} \&
  \bibinfo{author}{Sergii \surnamestart Strelchuk\surnameend}
  (\bibinfo{year}{2024}): \emph{\bibinfo{title}{Ultrafast hybrid
  fermion-to-qubit mapping}}.
\newblock {\slshape \bibinfo{journal}{Phys. Rev. B}} \bibinfo{volume}{109}, p.
  \bibinfo{pages}{115149}, \doi{10.1103/PhysRevB.109.115149}.
\newblock \urlprefix\url{https://link.aps.org/doi/10.1103/PhysRevB.109.115149}.

\bibitemdeclare{misc}{poorZXcalculusCompleteFiniteDimensional2024}
\bibitem{poorZXcalculusCompleteFiniteDimensional2024}
\bibinfo{author}{Boldizs{\'a}r \surnamestart Po{\'o}r\surnameend},
  \bibinfo{author}{Razin~A. \surnamestart Shaikh\surnameend} \&
  \bibinfo{author}{Quanlong \surnamestart Wang\surnameend}
  (\bibinfo{year}{2024}): \emph{\bibinfo{title}{{{ZX-calculus}} Is {{Complete}}
  for {{Finite-Dimensional Hilbert Spaces}}}}.
\newblock \eprint{2405.10896}.

\bibitemdeclare{inproceedings}{poorCompletenessArbitraryFinite2023}
\bibitem{poorCompletenessArbitraryFinite2023}
\bibinfo{author}{Boldizs{\'a}r \surnamestart Po{\'o}r\surnameend},
  \bibinfo{author}{Quanlong \surnamestart Wang\surnameend},
  \bibinfo{author}{Razin~A. \surnamestart Shaikh\surnameend},
  \bibinfo{author}{Lia \surnamestart Yeh\surnameend}, \bibinfo{author}{Richie
  \surnamestart Yeung\surnameend} \& \bibinfo{author}{Bob \surnamestart
  Coecke\surnameend} (\bibinfo{year}{2023}): \emph{\bibinfo{title}{Completeness
  for Arbitrary Finite Dimensions of {{ZXW-calculus}}, a Unifying Calculus}}.
\newblock In: {\slshape \bibinfo{booktitle}{2023 38th {{Annual ACM}}/{{IEEE
  Symposium}} on {{Logic}} in {{Computer Science}} ({{LICS}})}},
  \bibinfo{address}{Boston, MA, USA}, pp. \bibinfo{pages}{1--14},
  \doi{10.1109/LICS56636.2023.10175672}.
\newblock \eprint{2302.12135}.

\bibitemdeclare{misc}{rodatz2024distancepreserving}
\bibitem{rodatz2024distancepreserving}
\bibinfo{author}{Benjamin \surnamestart Rodatz\surnameend},
  \bibinfo{author}{Boldizsár \surnamestart Poór\surnameend} \&
  \bibinfo{author}{Aleks \surnamestart Kissinger\surnameend}
  (\bibinfo{year}{2024}): \emph{\bibinfo{title}{Floquetifying stabiliser codes
  with distance-preserving rewrites}}.
\newblock \eprint{2410.17240}.

\bibitemdeclare{article}{Seeley_2012}
\bibitem{Seeley_2012}
\bibinfo{author}{Jacob~T. \surnamestart Seeley\surnameend},
  \bibinfo{author}{Martin~J. \surnamestart Richard\surnameend} \&
  \bibinfo{author}{Peter~J. \surnamestart Love\surnameend}
  (\bibinfo{year}{2012}): \emph{\bibinfo{title}{The Bravyi-Kitaev
  transformation for quantum computation of electronic structure}}.
\newblock {\slshape \bibinfo{journal}{The Journal of Chemical Physics}}
  \bibinfo{volume}{137}(\bibinfo{number}{22}), \doi{10.1063/1.4768229}.
\newblock \urlprefix\url{http://dx.doi.org/10.1063/1.4768229}.

\bibitemdeclare{article}{Setia2019superfast}
\bibitem{Setia2019superfast}
\bibinfo{author}{Kanav \surnamestart Setia\surnameend}, \bibinfo{author}{Sergey
  \surnamestart Bravyi\surnameend}, \bibinfo{author}{Antonio \surnamestart
  Mezzacapo\surnameend} \& \bibinfo{author}{James~D. \surnamestart
  Whitfield\surnameend} (\bibinfo{year}{2019}): \emph{\bibinfo{title}{Superfast
  encodings for fermionic quantum simulation}}.
\newblock {\slshape \bibinfo{journal}{Physical Review Research}}
  \bibinfo{volume}{1}(\bibinfo{number}{3}),
  \doi{10.1103/physrevresearch.1.033033}.
\newblock \urlprefix\url{http://dx.doi.org/10.1103/PhysRevResearch.1.033033}.

\bibitemdeclare{misc}{shaikhCategoricalSemanticsFeynman2022}
\bibitem{shaikhCategoricalSemanticsFeynman2022}
\bibinfo{author}{Razin~A. \surnamestart Shaikh\surnameend} \&
  \bibinfo{author}{Stefano \surnamestart Gogioso\surnameend}
  (\bibinfo{year}{2022}): \emph{\bibinfo{title}{Categorical {{Semantics}} for
  {{Feynman Diagrams}}}}.
\newblock \eprint{2205.00466}.

\bibitemdeclare{misc}{shaikhHowSumExponentiate2022}
\bibitem{shaikhHowSumExponentiate2022}
\bibinfo{author}{Razin~A. \surnamestart Shaikh\surnameend},
  \bibinfo{author}{Quanlong \surnamestart Wang\surnameend} \&
  \bibinfo{author}{Richie \surnamestart Yeung\surnameend}
  (\bibinfo{year}{2022}): \emph{\bibinfo{title}{How to Sum and Exponentiate
  {{Hamiltonians}} in {{ZXW}} Calculus}}.
\newblock \eprint{2212.04462}.

\bibitemdeclare{misc}{shaikhFockedupZX2024}
\bibitem{shaikhFockedupZX2024}
\bibinfo{author}{Razin~A. \surnamestart Shaikh\surnameend},
  \bibinfo{author}{Lia \surnamestart Yeh\surnameend} \&
  \bibinfo{author}{Stefano \surnamestart Gogioso\surnameend}
  (\bibinfo{year}{2024}): \emph{\bibinfo{title}{The {{Focked-up ZX Calculus}}:
  {{Picturing Continuous-Variable Quantum Computation}}}}.
\newblock \eprint{2406.02905}.

\bibitemdeclare{inproceedings}{Somma2003qsimboson}
\bibitem{Somma2003qsimboson}
\bibinfo{author}{Rolando~D. \surnamestart Somma\surnameend},
  \bibinfo{author}{Gerardo \surnamestart Ortiz\surnameend},
  \bibinfo{author}{Emanuel~H. \surnamestart Knill\surnameend} \&
  \bibinfo{author}{James \surnamestart Gubernatis\surnameend}
  (\bibinfo{year}{2003}): \emph{\bibinfo{title}{Quantum simulations of physics
  problems}}.
\newblock In \bibinfo{editor}{Eric \surnamestart Donkor\surnameend},
  \bibinfo{editor}{Andrew~R. \surnamestart Pirich\surnameend} \&
  \bibinfo{editor}{Howard~E. \surnamestart Brandt\surnameend}, editors:
  {\slshape \bibinfo{booktitle}{Quantum Information and Computation}},
  \bibinfo{publisher}{SPIE}, \doi{10.1117/12.487249}.

\bibitemdeclare{article}{Steudtner2019AQMs}
\bibitem{Steudtner2019AQMs}
\bibinfo{author}{Mark \surnamestart Steudtner\surnameend} \&
  \bibinfo{author}{Stephanie \surnamestart Wehner\surnameend}
  (\bibinfo{year}{2019}): \emph{\bibinfo{title}{Quantum codes for quantum
  simulation of fermions on a square lattice of qubits}}.
\newblock {\slshape \bibinfo{journal}{Phys. Rev. A}} \bibinfo{volume}{99}, p.
  \bibinfo{pages}{022308}, \doi{10.1103/PhysRevA.99.022308}.
\newblock \urlprefix\url{https://link.aps.org/doi/10.1103/PhysRevA.99.022308}.

\bibitemdeclare{article}{Sutcliffe2024ZXCutting}
\bibitem{Sutcliffe2024ZXCutting}
\bibinfo{author}{Matthew \surnamestart Sutcliffe\surnameend} \&
  \bibinfo{author}{Aleks \surnamestart Kissinger\surnameend}
  (\bibinfo{year}{2024}): \emph{\bibinfo{title}{Procedurally Optimised
  ZX-Diagram Cutting for Efficient T-Decomposition in Classical Simulation}}.
\newblock {\slshape \bibinfo{journal}{Electronic Proceedings in Theoretical
  Computer Science}} \bibinfo{volume}{406}, pp. \bibinfo{pages}{63--78},
  \doi{10.4204/eptcs.406.3}.

\bibitemdeclare{article}{VerstraeteCirac2005local}
\bibitem{VerstraeteCirac2005local}
\bibinfo{author}{F~\surnamestart Verstraete\surnameend} \& \bibinfo{author}{J~I
  \surnamestart Cirac\surnameend} (\bibinfo{year}{2005}):
  \emph{\bibinfo{title}{Mapping local Hamiltonians of fermions to local
  Hamiltonians of spins}}.
\newblock {\slshape \bibinfo{journal}{Journal of Statistical Mechanics: Theory
  and Experiment}} \bibinfo{volume}{2005}(\bibinfo{number}{09}), p.
  \bibinfo{pages}{P09012–P09012}, \doi{10.1088/1742-5468/2005/09/p09012}.
\newblock \urlprefix\url{http://dx.doi.org/10.1088/1742-5468/2005/09/P09012}.

\bibitemdeclare{inproceedings}{vilmartNearMinimalAxiomatisationZXCalculus2019}
\bibitem{vilmartNearMinimalAxiomatisationZXCalculus2019}
\bibinfo{author}{Renaud \surnamestart Vilmart\surnameend}
  (\bibinfo{year}{2019}): \emph{\bibinfo{title}{A {{Near-Minimal
  Axiomatisation}} of {{ZX-Calculus}} for {{Pure Qubit Quantum Mechanics}}}}.
\newblock In: {\slshape \bibinfo{booktitle}{2019 34th {{Annual ACM}}/{{IEEE
  Symposium}} on {{Logic}} in {{Computer Science}} ({{LICS}})}}, pp.
  \bibinfo{pages}{1--10}, \doi{10.1109/LICS.2019.8785765}.
\newblock \eprint{1812.09114}.

\bibitemdeclare{article}{vlasovCliffordAlgebrasSpin2022}
\bibitem{vlasovCliffordAlgebrasSpin2022}
\bibinfo{author}{Alexander~Yurievich \surnamestart Vlasov\surnameend}
  (\bibinfo{year}{2022}): \emph{\bibinfo{title}{Clifford {{Algebras}}, {{Spin
  Groups}} and {{Qubit Trees}}}}.
\newblock {\slshape \bibinfo{journal}{Quanta}}
  \bibinfo{volume}{11}(\bibinfo{number}{1}), pp. \bibinfo{pages}{97--114},
  \doi{10.12743/quanta.v11i1.199}.

\bibitemdeclare{misc}{wangCompletenessQufiniteZXW2024}
\bibitem{wangCompletenessQufiniteZXW2024}
\bibinfo{author}{Quanlong \surnamestart Wang\surnameend},
  \bibinfo{author}{Boldizs{\'a}r \surnamestart Po{\'o}r\surnameend} \&
  \bibinfo{author}{Razin~A. \surnamestart Shaikh\surnameend}
  (\bibinfo{year}{2024}): \emph{\bibinfo{title}{Completeness of Qufinite
  {{ZXW}} Calculus, a Graphical Language for Finite-Dimensional Quantum
  Theory}}.
\newblock \eprint{2309.13014}.

\bibitemdeclare{article}{Whitfield2016local}
\bibitem{Whitfield2016local}
\bibinfo{author}{James~D. \surnamestart Whitfield\surnameend},
  \bibinfo{author}{Vojtěch \surnamestart Havlíček\surnameend} \&
  \bibinfo{author}{Matthias \surnamestart Troyer\surnameend}
  (\bibinfo{year}{2016}): \emph{\bibinfo{title}{Local spin operators for
  fermion simulations}}.
\newblock {\slshape \bibinfo{journal}{Physical Review A}}
  \bibinfo{volume}{94}(\bibinfo{number}{3}), \doi{10.1103/physreva.94.030301}.
\newblock \urlprefix\url{http://dx.doi.org/10.1103/PhysRevA.94.030301}.

\end{thebibliography}
